\documentclass[12pt,preprint]{aastex}

\shorttitle{Cluster Structure Evolution}

\shortauthors{Jeltema et al.}

\begin{document}

\title{The Evolution of Structure in X-ray Clusters of Galaxies}

\author{Tesla E. Jeltema\footnote{current address: Carnegie Observatories; 813 Santa Barbara St; Pasadena, CA 91101}, Claude R. Canizares, Mark W. Bautz}

\affil{Center for Space Research and Department of Physics}
\affil{Massachusetts Institute of Technology}
\affil{70 Vassar Street, Cambridge, MA 02139}

\email{tesla@ociw.edu}

\author{David A. Buote}

\affil{Department of Physics and Astronomy}
\affil{University of California at Irvine}
\affil{4129 Frederick Reines Hall, Irvine, CA 92697}

\begin{abstract}
Using Chandra archival data, we quantify the evolution of cluster morphology with redshift.  Clusters form and grow through mergers with other clusters and groups, and the amount of substructure in clusters in the present epoch and how quickly it evolves with redshift depend on the underlying cosmology.  Our sample includes 40 X-ray selected, luminous clusters from the Chandra archive, and we quantify cluster morphology using the power ratio method (Buote \& Tsai 1995).  The power ratios are constructed from the moments of the X-ray surface brightness and are related to a cluster's dynamical state.  We find that, as expected qualitatively from hierarchical models of structure formation, high-redshift clusters have more substructure and are dynamically more active than low-redshift clusters.  Specifically, the clusters with $z>0.5$ have significantly higher average third and fourth order power ratios than the lower redshift clusters.  Of the power ratios, $P_3/P_0$ is the most unambiguous indicator of an asymmetric cluster structure, and the difference in $P_3/P_0$ between the two samples remains significant even when the effects of noise and other systematics are considered.  After correcting for noise, we apply a linear fit to $P_3/P_0$ versus redshift and find that the slope is greater than zero at better than 99\% confidence.  This observation of structure evolution indicates that dynamical state may be an important systematic effect in cluster studies seeking to constrain cosmology, and when calibrated against numerical simulations, structure evolution will itself provide interesting bounds on cosmological models.
\end{abstract}


\section{INTRODUCTION}

Clusters form and grow through mergers with other clusters and groups.  These mergers are observed as multiple peaks in the cluster density distribution or as disturbed cluster morphologies.  The cluster relaxation time is relatively short, on the order of a Gyr, and the fraction of unrelaxed clusters should reflect their formation rate.  The formation epoch of clusters depends on both $\Omega_m$ and $\Lambda$, so the amount of substructure in clusters in the present epoch and how quickly it evolves with redshift depend on these parameters.  In low density universes, clusters form earlier and will be on average more relaxed in the present epoch.  Clusters at high redshift, closer to the epoch of cluster formation, should be on average dynamically younger and show more structure.  In addition, mergers introduce systematic errors in other cosmological studies with clusters.  Mergers can lead to large deviations in cluster luminosity, temperature, and velocity dispersion (Rowley, Thomas, \& Kay 2004; Randall, Sarazin, \& Ricker 2002; Mathiesen \& Evrard 2001) and therefore errors in estimating cluster mass and gas mass fraction as well as deviations from equilibrium.  For these reasons, a method of measuring cluster dynamical state and an understanding of how it evolves are important.

A number of studies have been done on the structure of clusters at low-redshift.  One of the first systematic studies was conducted by Jones \& Forman (1992) who visually examined 208 clusters observed with the \textit{Einstein} X-ray satellite.  They separated these clusters into 6 morphological classes including SINGLE, ELLIPTICAL, OFFSET CENTER, PRIMARY WITH SMALL SECONDARY, DOUBLE, and COMPLEX.  They found that 40\% of their clusters fell into the latter five categories, and 22\% fell into the three categories exhibiting multiple peaks.  This study established that merging is common in clusters.  However, in order to test cosmology, a more quantitative measure of cluster structure and dynamical state is needed.  Methods to quantify structure have used both the X-ray surface brightness distribution and, at optical wavelengths, the distribution of cluster galaxies.  However, optical studies require a large number of galaxies (Dutta 1995), at least a few hundred, and are more susceptible to contamination from foreground and background objects.  The only method of observing clusters that directly probes cluster mass is lensing.  However, lensing is sensitive to the projection of mass along the line of sight and does not have good resolution outside the core of the cluster.  
For the above reasons, we chose to study X-ray cluster observations.

X-ray studies of cluster substructure use a number of different statistics (see Buote 2002 for a review).  For example, Mohr et al. (1995) measured centroid variation, axial ratio or ellipticity, orientation, and radial fall-off for a sample 65 clusters.  Several other studies have also used ellipticity (Gomez et al. 1997; Kolokotronis et al. 2001; Melott, Chambers, \& Miller 2001; Plionis 2002); however, ellipticity is not a clear indicator of morphology.  Relaxed systems can be elliptical, and substructure can be distributed symmetrically.  Centroid variation is a better method in which the cluster centroid is calculated in a number of circular annuli of increasing radius.  The emission weighted variation of this centroid is a measure of cluster structure (Mohr, Fabricant, \& Geller 1993; Mohr et al. 1995; Gomez et al. 1997; Rizza et al. 1998; Kolokotronis et al. 2001).  This method is most sensitive to equal mass double clusters.  

Another method developed by Buote \& Tsai (1995, 1996) to study a sample of 59 clusters observed with \textit{ROSAT} is the power ratio method.  The power ratios are constructed from the moments of the X-ray surface brightness.  This method has the advantages that it is both related to cluster dynamical state, and it is capable of distinguishing a large range of cluster morphologies.  For our study, we chose to use the power ratio method, and we describe it in more detail in section 3.

Richstone, Loeb, \& Turner (1992) performed the first theoretical study of the relationship of substructure to cosmology.  In their analytical calculations, they assumed that substructure is wiped out in a cluster crossing time, and they calculated the fraction of clusters in the spherical growth approximation which formed within the last crossing time as a function of $\Omega_m$ and $\Lambda$.  They find that this fraction primarily depends on $\Omega_m$, and they estimated that $\Omega_m \ge 0.5$ based on estimates of the frequency of substructure in low-redshift clusters (Jones \& Forman 1992).  This method, like the observational study of Jones \& Forman (1992), predicts the ill-defined fraction of clusters with substructure.  A more recent semi-analytical approach was developed by Buote (1998).  He assumes that the amount of substructure depends on the amount of mass accreted by a cluster over a relaxation timescale and relates the fractional accreted mass to the power ratios.

Although these semi-analytic methods give an indication of the expected evolution of cluster substructure and its dependence on cosmological parameters, perhaps the best method of constraining cosmological models is through the comparison to cluster simulations.  Numerical simulations show that both the centroid (or center-of-mass) shift and the power ratios are capable of distinguishing cosmological models (Evrard et al. 1993; Jing et al 1995; Dutta 1995; Crone, Evrard, \& Richstone 1996; Buote \& Xu 1997; Thomas et al. 1998; Valdarnini, Ghizzardi, \& Bonometto 1999; Suwa et al. 2003).  Unfortunately, comparison of the different observational studies to simulations have led to contradictory conclusions.  Mohr et al. (1995) found that their centroid shifts were consistent with $\Omega_m = 1$, and Buote \& Xu (1997) find that the power ratios of their \textit{ROSAT} clusters indicate an $\Omega_m<1$ universe.  Both of these studies have weaknesses.  Buote \& Xu (1997) used dark matter only simulations and approximated the power ratios of the X-ray surface brightness as the power ratios of the dark matter density squared.  Mohr et al. (1995) used simulations which incorporate the cluster gas, but which only included eight clusters.  In addition, these hydrodynamic simulations have poor force resolution for the gas.  

Cluster structure has also been examined in two more recent sets of hydrodynamic simulations.  Valdarnini et al. (1999) compute power ratios for clusters formed in three cosmological models: flat CDM, $\Lambda$CDM with $\Lambda = 0.7$, and CHDM with $\Omega_h = 0.2$ and one species of massive neutrino.  For each model, they simulated 40 clusters and compared the results to the \textit{ROSAT} sample studied by Buote \& Tsai (1996). 
They find that the $\Lambda$CDM model is inconsistent with the data, but neither CDM or CHDM are ruled out.  However, these simulations used $\sigma_8=1.1$.  This value of $\sigma_8$ is fairly high and may cause the disagreement between the $\Lambda$CDM model and the data.  In addition, these simulations neglect the effect of the tidal field at large scales.  Suwa et al. (2003) compare simulated clusters in a $\Lambda$CDM and an OCDM cosmology, at both $z=0$ and $z=0.5$, using several methods of quantifying structure.  They find that axial ratio and cluster clumpiness are not successful at distinguishing the two models.  However, the center shifts and the power ratios of both the surface brightness and the projected mass density are able to discriminate between the models at $z=0$.  The power ratios of the surface brightness are also successful at $z=0.5$.  They restrict themselves to comparing the ability of different statistical indicators to distinguish cosmologies and do not compare to observations
, but their results do show that cluster structure can potentially constrain $\Lambda$ or a time dependent vacuum energy.  

Obviously this situation needs further examination.  In a future study, we will examine cluster morphology and its evolution in simulated clusters from two independent hydrodynamic simulations.  The current work will focus on the observational side.  Specifically, we seek to place an additional constraint on cosmological models by examining the evolution of cluster structure with redshift.

All of the observational studies described above are limited to clusters with redshifts less than approximately 0.3.  Until recently, the number of clusters known with $z>0.5$ has been limited to a few.  However, recent surveys, notably the many \textit{ROSAT} surveys, have increased this number by an order of magnitude (Rosati et al. 1998; Perlman et al. 2002; Gioia et al. 2003; Vikhlinin et al. 1998).  It is now possible to study the evolution of cluster structure out to $z\sim1$.  Using a sample of 40 clusters observed with the \textit{Chandra} X-ray Observatory we show that the amount of substructure in clusters increases with redshift, as expected qualitatively from hierarchical models of structure formation.  
This paper is organized as follows:  In section 2, we describe our sample selection.  In \S3, we give a more detailed description of the power ratio method.  In \S4, we describe the data reduction and the calculation of uncertainties, and in section 5 we give our results.  Finally, we discuss the systematic effects which could influence our results, and we give our conclusions.  We assume a cosmology of $H_0=70h_{70}$ km s$^{-1}$ Mpc$^{-1}$, $\Omega_m=0.3$, and $\Lambda = 0.7$ throughout.

\section{ SAMPLE }

For this project, we used data from the \textit{Chandra} archive, which allowed us to select clusters over a large redshift range with observations of sufficient depth.  In addition, \textit{Chandra}'s superb resolution aids in the identification and exclusion of point sources from the analysis.  A lower limit of $z=0.1$ was placed on the redshift to ensure that a reasonable area of each cluster would be visible on a \textit{Chandra} CCD.  To ensure that the sample was relatively unbiased, all clusters were selected from flux-limited X-ray surveys.  The majority of the sample came from the \textit{Einstein} Medium Sensitivity Survey (EMSS; Gioia \& Luppino 1994) and the \textit{ROSAT} Brightest Cluster Sample (BCS; Ebeling et al. 1998).  Clusters were also required to have a luminosity greater than $5 \times 10^{44}$ ergs s$^{-1}$, as listed in those catalogs.  Additional high-redshift clusters were selected from recent \textit{ROSAT} surveys including the \textit{ROSAT} Deep Cluster Survey (RDCS), the Wide Angle \textit{ROSAT} Pointed Survey (WARPS), the \textit{ROSAT} North Ecliptic Pole Survey (NEP), and the 160 deg$^2$ survey (Rosati et al. 1998; Perlman et al. 2002; Gioia et al. 2003; Vikhlinin et al. 1998).  The added clusters were required to come from X-ray flux-limited surveys, have high-redshifts ($z>0.5$), be luminous, and have observations of sufficient depth to allow structure analysis.  They represent all such clusters with published redshifts available in the archive at the time of sample selection and many of the highest-redshift, highest-luminosity clusters discovered in the above surveys at that time.  The requirement of sufficient depth limited us to clusters with $z<0.9$, but did not exclude any clusters with lower redshifts.  The resulting sample contains 40 clusters with redshifts between 0.11 and 0.89.  The initial sample contained about 50 clusters; however, several clusters were removed due to complications in the data reduction such as a high soft X-ray background flux or background flares.

The clusters along with their redshifts, observation IDs, clean exposure times, and luminosities within a radius of 0.5 Mpc are listed in Table 1.  These luminosities were estimated from the \textit{Chandra} observations in a 0.3-7.0 keV band and using a Raymond-Smith thermal plasma model with $N_H=3\times10^{20}$ atoms cm$^{-2}$, $kT=5$ keV, and an abundance of 0.3 relative to the abundances of Anders \& Grevesse (1989).  Changes in the column density and temperature did not have a large effect on the luminosities.  The \textit{Chandra} luminosities range from $2.0\times10^{44}$ ergs s$^{-1}$ to $2.3\times10^{45}$ ergs s$^{-1}$.

In the following analysis, the clusters were divided into two samples.  The high-redshift sample contains the 14 clusters with $z>0.5$, and the low-redshift sample contains the 26 clusters with $z<0.5$.  The two samples have average redshifts of 0.24 and 0.71, respectively.  The average luminosities of the two samples are $7.4\times10^{44}$ ergs s$^{-1}$ for the high-redshift sample and $9.1\times10^{44}$ ergs s$^{-1}$ for the low-redshift sample.  As discussed later, this small difference in average luminosity does not effect our results.

\placetable{tbl-1}

\section{ POWER RATIO METHOD }
Here we present the power ratio method as developed by Buote \& Tsai (1995, 1996).  The power ratios are capable of distinguishing a large range of cluster morphologies.  Essentially, this method entails calculating the multipole moments of the X-ray surface brightness in a circular aperture centered on the cluster's centroid.  The moments, $a_m$ and $b_m$ given below, are sensitive to asymmetries in the surface brightness distribution and are, therefore, sensitive to substructure. For example, for a perfectly round cluster the $m>0$ moments are all zero, for a perfectly elliptical cluster or a perfectly equal mass, symmetric double cluster the odd moments are zero but the even moments are not, and for all of the other Jones \& Forman (1992) classifications the moments are all non-zero.  

The physical motivation for this method is that it is related to the multipole expansion of the two-dimensional gravitational potential.  Large potential fluctuations drive violent relaxation, and therefore, the power ratios are related to a cluster's dynamical state.  The multipole expansion of the two-dimensional gravitational potential is
\begin{equation}
\Psi(R,\phi) = -2Ga_0\ln\left({1 \over R}\right) -2G
\sum^{\infty}_{m=1} {1\over m R^m}\left(a_m\cos m\phi + b_m\sin
m\phi\right). \label{eqn.multipole}
\end{equation}
and the moments $a_m$ and $b_m$ are
\begin{eqnarray}
a_m(R) & = & \int_{R^{\prime}\le R} \Sigma(\vec x^{\prime})
\left(R^{\prime}\right)^m \cos m\phi^{\prime} d^2x^{\prime}, \nonumber \\
b_m(R) & = & \int_{R^{\prime}\le R} \Sigma(\vec x^{\prime})
\left(R^{\prime}\right)^m \sin m\phi^{\prime} d^2x^{\prime}, \nonumber
\end{eqnarray}
where $\vec x^{\prime} = (R^{\prime},\phi^{\prime})$ and $\Sigma$ is the surface mass density.  In the case of X-ray studies, X-ray surface brightness replaces surface mass density in the calculation of the power ratios.  
 X-ray surface brightness is proportional to the gas density squared and generally shows the same qualitative structure as the projected mass density, allowing a similar quantitative classification of clusters.  

The powers are formed by integrating the magnitude of $\Psi_m$, the \textit{m}th term in the multipole expansion of the potential given in equation (1), over a circle of radius $R$,
\begin{equation}
P_m(R)={1 \over 2\pi}\int^{2\pi}_0\Psi_m(R, \phi)\Psi_m(R, \phi)d\phi.
\end{equation}
Ignoring factors of $2G$, this gives
\begin{equation}
P_0=\left[a_0\ln\left(R\right)\right]^2 \nonumber
\end{equation}
\begin{equation}
P_m={1\over 2m^2 R^{2m}}\left( a^2_m + b^2_m\right). \nonumber
\end{equation}
Rather than using the powers themselves, we divide by $P_0$ to normalize out flux forming the power ratios, $P_m/P_0$.  

Here we consider only the observable two-dimensional cluster properties.  As we cannot know a cluster's three-dimensional structure, there is a degeneracy in the interpretation of the dynamical state of any individual cluster due to projection effects.  Since we are only concerned with relatively large-scale, cosmologically significant structure, it should be rare that a very disturbed cluster appears very relaxed in projection.  The power ratios essentially provide a lower limit on a cluster's true structure.  For a reasonable sample size, projection will not have a large effect.  In addition, cosmological constraints placed through the comparison of observed clusters to numerical simulations will be valid, because the power ratios can be calculated in a consistent way from projection of the simulated clusters.

For each cluster in our sample, we calculate $P_2/P_0$, $P_3/P_0$, and $P_4/P_0$ in a circular aperture centered on the centroid of cluster emission.  $P_1$ vanishes with the origin at the centroid, and the higher order terms are sensitive to successively smaller scale structures which are both more affected by noise and less cosmologically significant.  The results given here are for an aperture radius with a physical size of 0.5 Mpc.  The choice of aperture size will be discussed in more detail in the next section.

\section{ DATA REDUCTION AND UNCERTAINTIES }

\subsection{ Image Preparation }

For the most part, the data were prepared using the standard data processing, including updating the gain map, applying the pixel and PHA randomizations, and filtering on \textit{ASCA} grades 0, 2, 3, 4, and 6 and a status of zero\footnote{\textit{Chandra} Proposers' Observatory Guide http://cxc.harvard.edu/proposer/POG/, section ``ACIS''}.  In the case of ACIS-I observations performed in VFAINT (VF) mode, the additional background cleaning for VF mode data was also applied.  Events are graded to distinguish good X-ray events from cosmic rays.  In the standard data processing, events are graded based on the pulse height in a 3x3 pixel island; in VF mode, the data includes the pixel values in a 5x5 island, allowing the extra pixels to be used to flag bad events.  This extra cleaning was not applied to VF mode ACIS-S observations or observations at a focal plane temperature of $-110^{\circ}$C due to the lack of corresponding background data (see section 4.2).  We also did not apply the charge transfer inefficiency (CTI) correction.  This correction was released half-way through our analysis, and it was not released for all chips and all time periods.  As a test, we applied the CTI correction to a few of the clusters in our sample and found that it had virtually no effect on the flux in the energy band we use.  

The energy range was restricted to the 0.3-7.0 keV band.  The data were then filtered to exclude time periods with background flares using the script \textit{lc\underline{ }clean}.  The filtering excluded time periods when the count rate, excluding point sources and cluster emission, in the 0.3-10 keV band was not within 20\% of the quiescent rate. This filtering matches the filtering applied to the background fields we use\footnote{See http://cxc.harvard.edu/cal/Acis/Cal\underline{ }prods/bkgrnd/acisbg/COOKBOOK and the note about versions in the next section.}.  For ACIS-S observations, flares were detected on S3 only when the cluster emission did not cover more than 20\% of the chip; otherwise, flares were detected using the other back-illuminated CCD, S1.  Observations with high background rates compared to the background fields were removed from the sample.  We then normalized the cluster images by a map of the exposure.  Exposure maps were weighted according to a Raymond-Smith spectrum with $N_H=3\times10^{20}$ atoms cm$^{-2}$, $kT=5$ keV, and an abundance of 0.3 relative to solar.  

The last step in the preparation of cluster images was the removal of point sources.  We detected point sources using the CIAO routine \textit{wavdetect} with the significance threshold set to give approximately one false detection per cluster image and with wavelet scales of 1, 2, 4, 8, and 16 pixels.  Elliptical background regions were defined around each source such that they contained at least five background counts.  We adjusted source and background regions by hand to ensure that they did not overlap other sources or extend off the image.  In some cases, sources containing only a few counts that did not appear to be real were excluded, and the regions around bright sources were expanded to include more of the counts in the wings of the PSF. Finally, we removed the sources and filled the source regions using the CIAO tool \textit{dmfilth}.  This tool fits a polynomial to each background region, and it computes pixel values within the source region according to this fit.  For many clusters, wavdetect found a source at the center of the cluster which we did not remove.  In most cases, this source was simply the peak of the surface brightness distribution rather than a central X-ray point source. In all cases, the flux of the ``source'' was small compared to the cluster flux.  We computed the power ratios for several clusters both with and without the central ``source'' and found that they did not change significantly.

For one cluster, MS2053.7-0449, we merged two observations.  Several of the clusters in our sample had multiple observations in the archive, but in the other cases these observations were made at different focal plane temperatures or with different detectors.  Under these circumstances we chose to use only the longest observation.  Both observations of MS2053.7-0449 were prepared separately according to the steps described above.  The two observations were aligned by hand using six bright X-ray point sources that appear in both images.  After the exposure correction, we merged the two images.  Point sources were then detected and filled in the combined image.

\subsection{ Background }

To account for the X-ray background, we chose to use the ACIS ``blank-sky'' data sets\footnote{See the web pages http://cxc.harvard.edu/contrib/maxim/bg/index.html and http://cxc.harvard.edu/cal/Acis/Cal\underline{ }prods/bkgrnd/acisbg/COOKBOOK.  In order to maintain consistency throughout our analysis and between different focal plane temperatures and CCDs, we have used the version of the background files released before CALDB v2.17.}.  The X-ray background consists of a combination of the cosmic X-ray background, including two soft thermal components arising from the Galactic halo and power-law component arising from unresolved AGN, and unrejected particle background.  The blank-sky data sets are a combination of several relatively source free observations with any obvious sources removed.  An observation specific local background was not used because it is difficult to find an area of the detector free of cluster emission for the low-redshift clusters.  In addition, the blank-sky data sets allow us to extract a background from the same region of the CCD as the cluster emission which more accurately models spatial variation in the background.  The background files are divided into four time periods to account for changes in the background due to a changing focal plane temperature.  Our sample includes observations from the last three time periods (B, C, and D).  Period B includes the time period when the focal plane temperature was $-110^{\circ}$C, and periods C and D are for a focal plane temperature of $-120^{\circ}$C.  

The backgrounds were matched to the observation dates with a couple of exceptions.  The period D background files contained only observations performed in VF mode allowing these files to be filtered using the additional VF background cleaning, but the period B and C files did not.  For cluster observations performed in VF mode during period C, we used the period D background files.  Periods C and D have the same focal plane temperature, and the backgrounds are not very different.  For VF mode, period B cluster observations, we used the period B background files and did not perform the VF background cleaning.  The period C background files for the ACIS-S detector included observations in the North Polar Spur, and they have a high soft background rate.  Therefore, period D background files were also used for ACIS-S, period C observations.

To prepare background images for each cluster observation we took the following steps.  First, the gain file of the background was updated to match the gain file used on the corresponding cluster observation.  The background events were then reprojected using the cluster aspect file to match the coordinates of the observation.  We limited the energy range to 0.3-7.0 keV, and for the D, VF backgrounds we applied the VF filtering.  The background files have a much longer exposure time than the observations and must be normalized to match them.  We normalized the backgrounds by the ratio of the counts in the 10-12 keV band in the observation to the 10-12 keV band counts in the background.  These numbers were calculated after removing cluster and source regions from both the observation and the background, and they were calculated separately for each CCD in the image.  The 10-12 keV energy band is used because it is above the passband of the grazing incidence optics.  Therefore, it is relatively free of sky emission, and the count rate in this band is fairly constant between observations\footnote{http://cxc.harvard.edu/cal/Acis/Cal\underline{ }prods/bkgrnd/acisbg/COOKBOOK}.  We checked these normalizations against the ratio of the exposure times to make sure they were not too different.  Another check performed was a comparison of the soft band flux for the cluster pointings versus the pointings included in the background files using the \textit{ROSAT} All-Sky Survey (RASS) R4-R5 band data (Snowden et al. 1997).  This check is necessary to ensure that the background files will sufficiently match the background in the observations.  A few clusters located in regions of high RASS flux were excluded from the sample.

The last step in the preparation of the background images was to properly account for bad pixels and columns.  Pixels flagged as bad include ``hot'' pixels, node boundaries, pixels with bias errors, and pixels adjacent to bad pixels.  Individual pixels flagged as bad would have little effect on the power ratios; however, bad columns may be important.  Unfortunately, the bad pixel lists excluded from the background files do not match the observation bad pixel files.  For example, pixels and columns adjacent to bad pixels/columns were not excluded from the backgrounds but are excluded from the observations.  To account for this mismatch, we made two instrument only exposure maps (i.e. they do not include the effects of the mirror and quantum efficiency) for each background image, one with the background bad pixels and one with the observation bad pixels.  We then divided the background image by the background bad pixel exposure map and multiplied it by the observation bad pixel map.  Finally, the background image was exposure corrected using the full observation exposure map (including the mirror and QE).  For MS2053.7-0449, the background images for each observation were prepared separately and then added.

\subsection{ Power Ratio Calculation and Estimation of Uncertainties }

The power ratios were calculated in circular apertures for a range of aperture radii varying from physical sizes of 0.1 Mpc to 1.2 Mpc (or the largest aperture which completely fit in the area of the detector) in 0.1 Mpc intervals.  We assumed a cosmology of $H_0=70 h_{70}$ km s$^{-1}$ Mpc$^{-1}$, $\Omega_m=0.3$, and $\Lambda = 0.7$.  First, the centroids (where $P_1$ vanishes) were calculated for each aperture starting with the largest aperture that fit on the detector and iterating in to an aperture radius of 0.1 Mpc.  For each aperture, we then calculated $P_0$, $P_2$, $P_3$, and $P_4$.  In the calculation of the powers, the X-ray background was accounted for by calculating the moments ($a_0-a_4$ and $b_0-b_4$) separately for the cluster image and the background image and then subtracting the background moments from the observed moments.  We then used the net cluster moments to find the powers.  

Some of the high-redshift clusters have relatively small fluxes, so it was desirable to set cutoffs for the minimum acceptable number of cluster counts and the minimum signal-to-noise (S/N) level.  We, therefore, only considered apertures with more than 500 net counts and with a S/N in the annulus surrounding the next smaller aperture of greater than 3.0.  Below this signal-to-noise level the program often had difficulty locating the centroid.  We found that an aperture radius of 0.5 Mpc was the largest radius for which all of the high-redshift clusters had an acceptable S/N and which was small enough not to extend beyond the detector for any of the low-redshift clusters.  The following analysis and discussion will be limited to this aperture size.  The power ratios for a radius of 0.5 Mpc are listed in Table 2.

\placetable{tbl-2}

To estimate the uncertainty in the power ratios due to noise, we employed a Monte Carlo method (Buote \& Tsai 1996).  First, the exposure corrected cluster images were adaptively binned, using the program \textit{AdaptiveBin} developed by Sanders and Fabian (2001), to give a minimum of two counts per bin.  The counts in each bin are averaged over the pixels in that bin to retain the same pixellation as the original image.  This procedure removed almost all zero pixels, although a few remained because we required that only adjacent pixels be binned together.  To add back instrumental effects, the binned image was multiplied by the exposure map.  We then added Poisson noise by taking each pixel value as the mean for a Poisson distribution and then randomly selecting a new pixel value from that distribution.  Finally, the image was exposure corrected.  This process was repeated 100 times for each cluster creating 100 mock cluster observations.  A few of the clusters in our sample had background moments similar to the cluster moments.  For these clusters, the effects of noise in the background images become important, and we also created 100 mock background images.

We calculated the power ratios for each of the 100 mock cluster images, and the 90\% confidence limits were defined to be the fifth highest and fifth lowest ratios.  These confidence intervals are listed in Table 3.  In the case of MS2053.7-0449, we created 100 mock images for each of the two cluster observations and merged them before calculating the power ratios.  For three of the clusters in our sample, one of the observed power ratios falls below the uncertainties.  We expect that a few clusters will fall outside of their errors, because the errors are only 90\% limits.  In all three cases, the power ratios for the surrounding $R=0.4$ Mpc and $R=0.6$ Mpc aperture sizes are more reflective of the position of the uncertainties.

We did not include the effects of point sources in the error estimation because neither unresolved sources nor the details of the source filling have a large effect on the power ratios. We estimated the flux from unresolved sources for a high-redshift, low-exposure cluster, CLJ0152.7-1357, using the LogN-LogS relation from the Chandra Deep Field South (Campana et al. 2001; Tozzi et al. 2001).  This flux amounted to at most 13\% of the total cluster flux.  At this flux level, the distribution of unresolved X-ray sources would have to be very non-uniform to have a significant effect on the power ratios.  As far as the source removal and filling, the number of source counts removed from the 0.5 Mpc aperture only exceeded 10\% of the cluster counts in a few cases and was never greater than 25\% of the cluster counts.  The percentage of the counts added by the filling of the source regions was quite a bit smaller.  For three clusters, we experimented with the level of binning by trying both binning to give 25 counts per bin and not binning at all.  The binning criteria did not seem to have a large effect on the errors derived.  However, large amounts of binning tended to smooth out cluster structure and led to a shift in the derived errors toward lower values of the power ratios.

An additional systematic effect considered was the normalization of the background.  To estimate the size of the possible error in the background normalizations, we compared the source free 0.3-7.0 keV flux in the observations to the same flux in the normalized backgrounds.  We estimate that the background normalizations could be off by a factor between 0.9 and 1.2.  We, therefore, reran the error calculations using both a background of $0.9*$(background image) and $1.2*$(background image) to bound the possible effect on the power ratios.  For cluster observations where the 0.5 Mpc aperture fell on multiple CCDs, we renormalized each CCD by 0.9 and 1.2 separately creating $2^{\# of chips}$ background images, and we ran the error calculation for each of these backgrounds.  The systematic errors are listed in brackets in Table 3 next to the corresponding noise errors.  We defined the systematic error to be the difference between the average of the 100 power ratios calculated with the original background and the average of the 100 power ratios calculated with the renormalized background.  The renormalization of 0.9 creates a shift toward smaller power ratios, and the renormalization of 1.2 creates a shift toward larger power ratios.  In a couple of cases, both renormalizations caused a shift of the same sign ($P_4/P_0$ for RXJ0439.0+0715 and $P_3/P_0$ for MS0015.9+1609) and the corresponding error was set to zero.  For the eight clusters where the 0.5 Mpc aperture fell on multiple CCDs, we list the maximum offset in the average power ratios of the $2^{\# of chips}$ renormalized runs from the original background run as the systematic error.

\placetable{tbl-3}

\section{ RESULTS }

Figure 1 illustrates the ability of the power ratios to distinguish different cluster morphologies.  The central plot shows $P_2/P_0$ versus $P_3/P_0$ (the quadrupole ratio versus the octupole ratio) for the 40 clusters in our sample.  The different power ratios are sensitive to different types of structure and the correlations among them aid in differentiating cluster morphologies.  In this plot, one could imagine a roughly diagonal line with the most disturbed clusters appearing at the upper-right and the most relaxed clusters appearing at the lower-left.  Smoothed \textit{Chandra} images for six clusters are also shown with their power ratios indicated.  These images all have the same physical scale of 1.4 Mpc on a side, and while the images contain X-ray point sources, these sources were removed before calculating the power ratios.  Both the double cluster CLJ0152.7-1357 and the complex cluster V1121.0+2327 have high power ratios (upper-right), while RXJ0439.0+0520, a relatively round, relaxed cluster, has small power ratios.  In between these are two clusters with smaller scale substructure.  A1413, an elliptical cluster, has similar $P_3/P_0$ but higher $P_2/P_0$ than RXJ0439.0+0520 (odd multipoles are not sensitive to ellipticity).

\begin{figure}
\centering
\epsscale{0.95}
\plotone{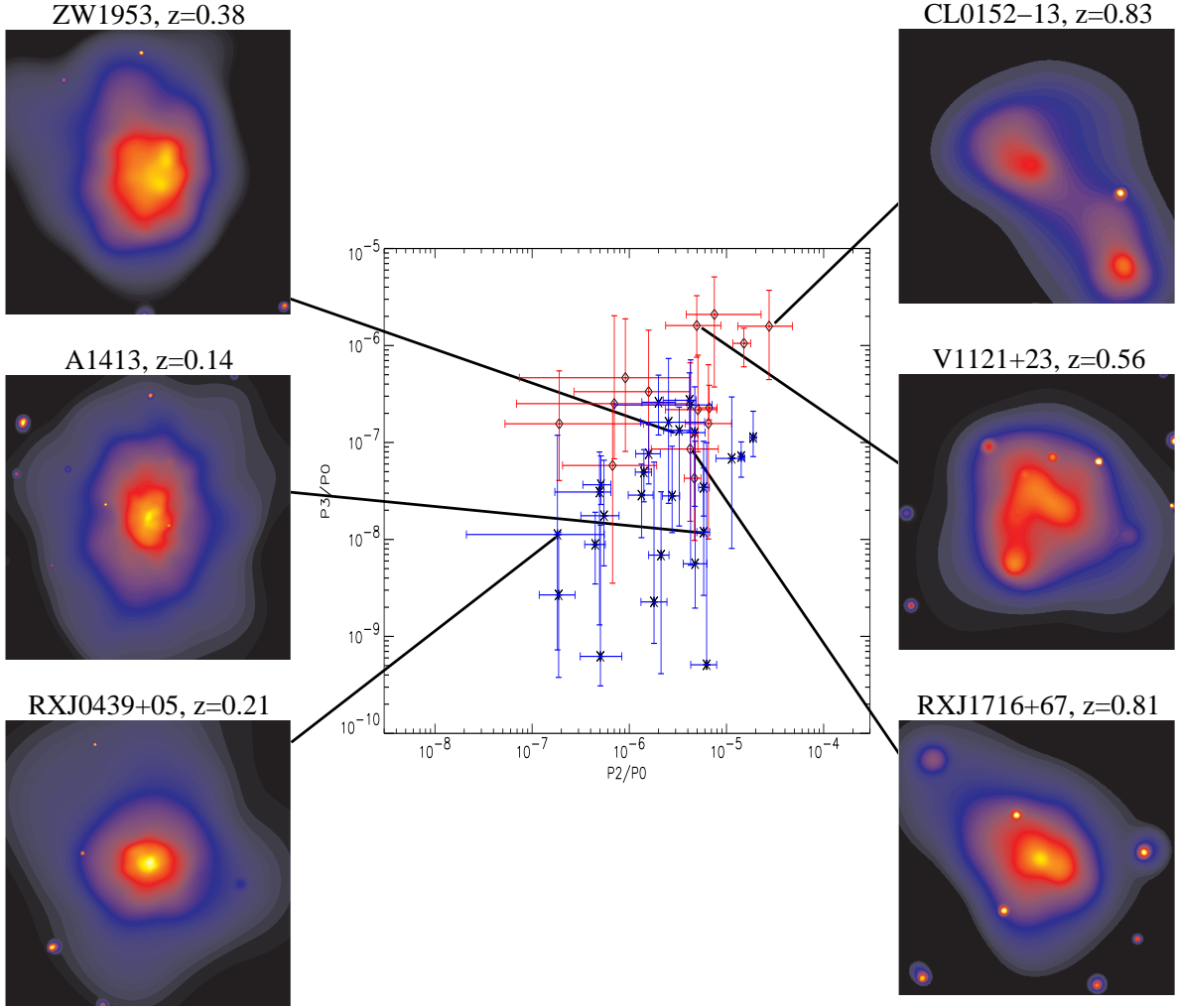}
\figcaption{ The central plot shows $P_2/P_0$ versus $P_3/P_0$.  Smoothed images for six clusters are shown with their power ratios indicated.  Both the double cluster CLJ0152.7-1357 and the complex cluster V1121.0+2327 have high power ratios (upper-right), while RXJ0439.0+0520, a relatively round, relaxed cluster, has small power ratios.  In between these are clusters with smaller scale substructure.  A1413, an elliptical cluster, has similar $P_3/P_0$ but higher $P_2/P_0$ than RXJ0439.0+0520 (odd multipoles are not sensitive to ellipticity).  Power ratios are computed in a 0.5 Mpc radius aperture.  High-redshift clusters ($z>0.5$) are plotted with diamonds and have red error bars.  Low-redshift clusters are plotted with asterisks and have blue error bars.  The images were adaptively smoothed using the CIAO routine \textit{csmooth}. }
\end{figure}

Figures 2-4 show the three possible projections of the power ratios.  Here the high-redshift clusters are plotted with diamonds and have red error bars.  The low-redshift clusters are plotted with asterisks and have blue error bars.  These error bars represent the noise only 90\% confidence limits.  It is apparent from these plots that the two samples have a similar distribution of $P_2/P_0$; however, the high-redshift clusters tend to have both higher $P_3/P_0$ and higher $P_4/P_0$ than the low-redshift clusters.  In particular, in the plot of $P_3/P_0$ versus $P_4/P_0$ the high-redshift clusters appear for the most part in the upper corner of the plot, while the low-redshift clusters tend toward the lower corner.  These results indicate that the high-redshift clusters have, on average, more substructure and are dynamically young compared to the low-redshift clusters.  The fact that $P_2/P_0$ does not distinguish the two samples could stem from its sensitivity to ellipticity.  As mentioned before, ellipticity is not a clear indicator of dynamical state.  In Figure 1, CLJ0152.7-1357 has significant ellipticity, but V1121.0+2327, which also shows significant substructure, is comparatively round.  In addition, A1413 is fairly elliptical but also fairly relaxed.  Ellipticity also contributes to $P_4/P_0$, but this ratio is more sensitive to smaller scale structure than $P_2/P_0$.  
$P_3/P_0$ best distinguishes the high and low-redshift clusters.  As an odd multipole term, this ratio is not sensitive to ellipticity, and a large $P_3/P_0$ is a clear indication of an asymmetric cluster structure.

\begin{figure}
\centering
\epsscale{0.7}
\plotone{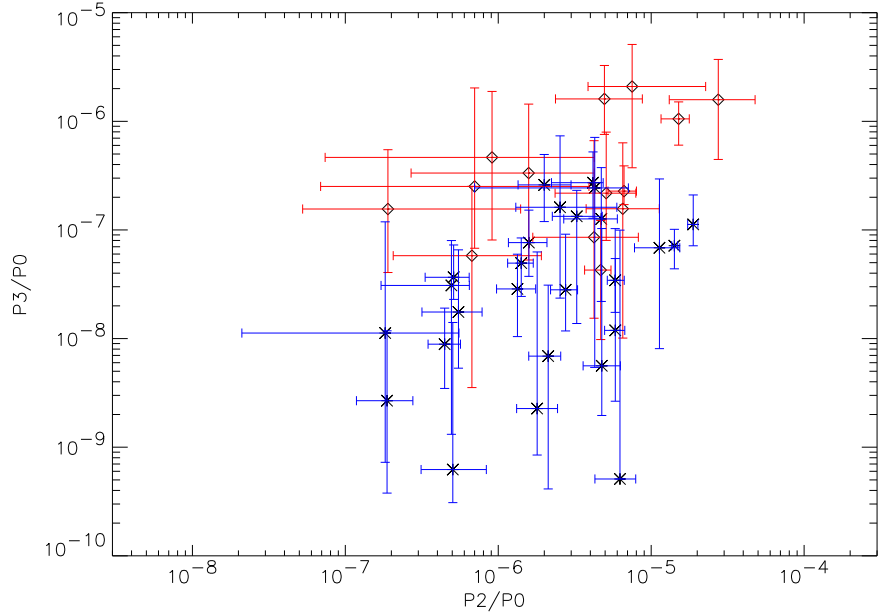}
\figcaption{ $P_2/P_0$ versus $P_3/P_0$.  Power ratios are computed in a 0.5 Mpc radius aperture.  High-redshift clusters ($z>0.5$) are plotted with diamonds and have red error bars.  Low-redshift clusters are plotted with asterisks and have blue error bars. }
\end{figure}

\begin{figure}
\centering
\epsscale{0.7}
\plotone{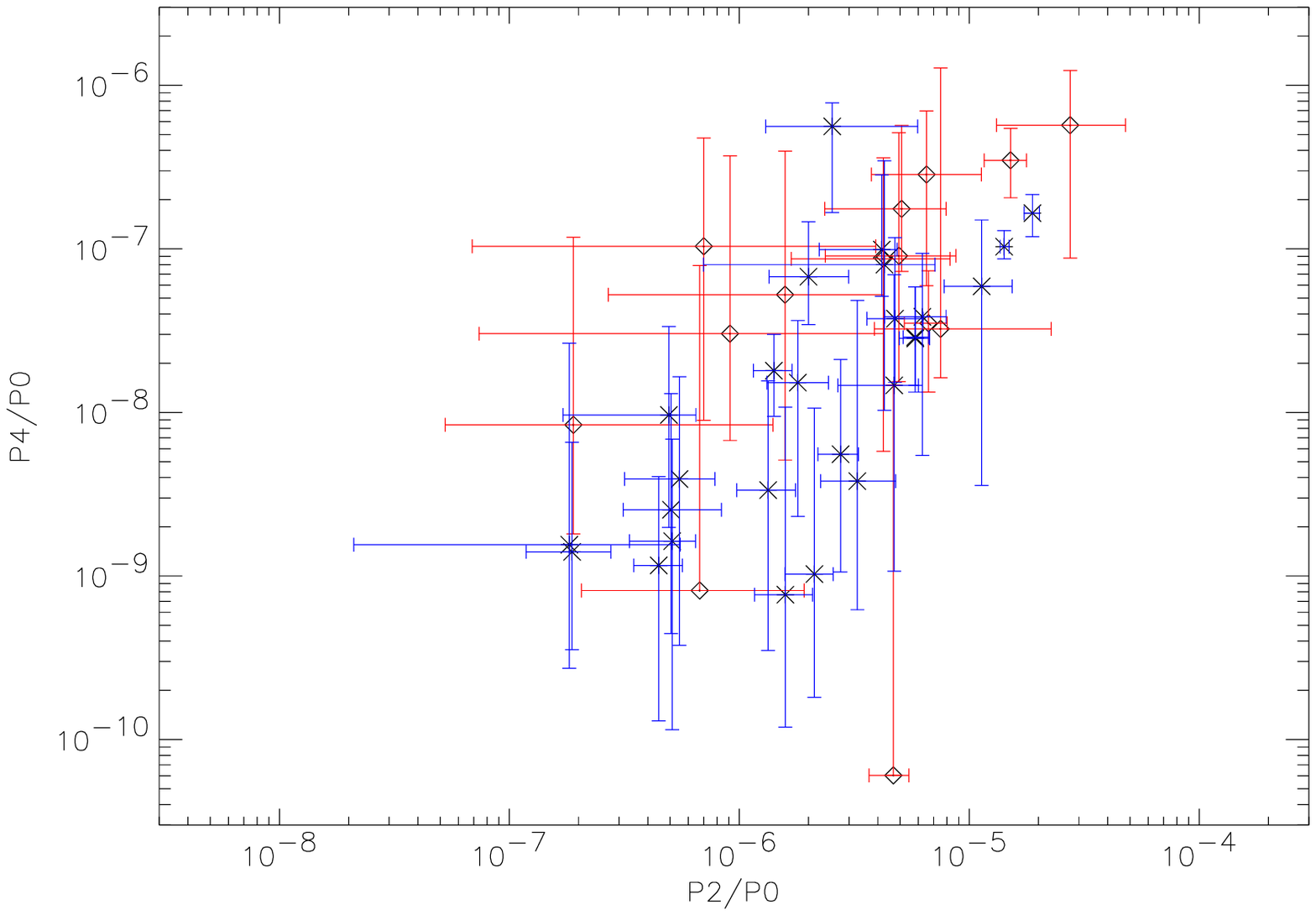}
\figcaption{ Same as Figure 2 for $P_2/P_0$ versus $P_4/P_0$. }
\end{figure}

\begin{figure}
\centering
\epsscale{0.7}
\plotone{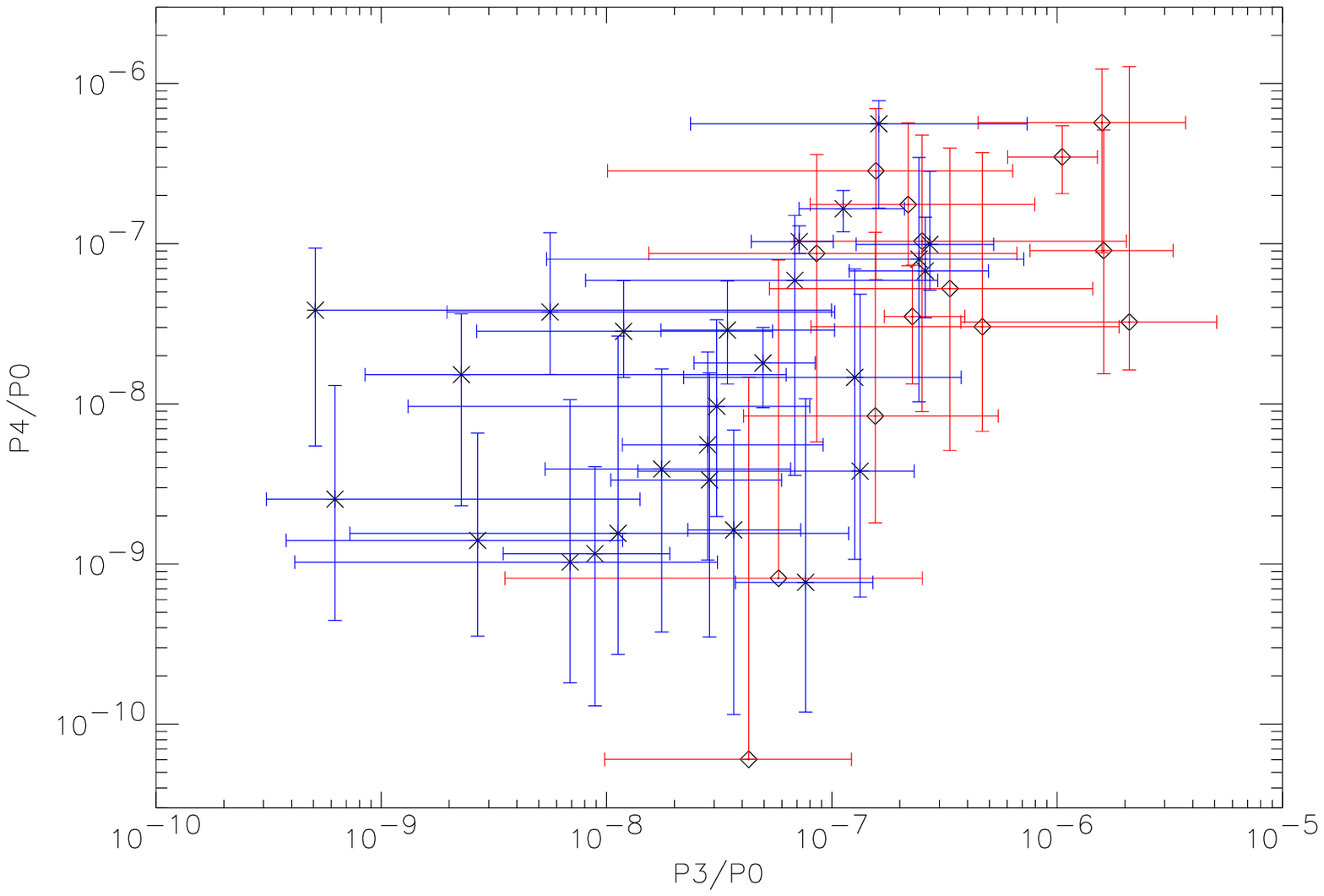}
\figcaption{ Same as Figure 2 for $P_3/P_0$ versus $P_4/P_0$. }
\end{figure}

We performed a number of tests to establish the statistical significance of the difference in power ratios between the high and low-redshift samples.  A Wilcoxon rank-sum test (e.g., Walpole \& Myers 1993) gives a probability of $4.6\times10^{-5}$ that the high and low-redshift clusters have the same mean $P_3/P_0$ and a probability of 0.025 for $P_4/P_0$.  A Kolmogorov-Smirnov test (e.g., Press et al. 1992, \S 14.3) also shows that the distributions of $P_3/P_0$ and $P_4/P_0$ differ significantly for the high and low-redshift samples, giving probabilities of 0.00064 and 0.041 for $P_3/P_0$ and $P_4/P_0$, respectively.  These tests do not show a significant difference in the distributions of $P_2/P_0$.

Unfortunately, the above tests do not include the uncertainties in the power ratios.  However, we have the results of the Monte Carlo simulations from which we can resample the power ratios for each cluster.  To account for both the noise and systematic errors we combined the results of the error calculations for the three background normalizations, giving 300 sets of power ratios for each cluster.  We then randomly selected $P_3/P_0$ and $P_4/P_0$ from these 300 for each cluster and reran the rank-sum and Kolmogorov-Smirnov (KS) tests.  This process was repeated 1000 times.  For $P_3/P_0$, the rank-sum test never gives a probability higher than 0.018.  The KS test gives an average probability of 0.0023 and only gives a probability greater than 0.05 for 5 out of 1000 runs.  The result that the high-redshift clusters generally have higher $P_3/P_0$ is therefore highly significant.  For $P_4/P_0$, the average rank-sum probability is 0.0082 with 28 of 1000 runs yielding a probability greater than 0.05.  For the KS test, these numbers are 0.037 and 204 out of 1000, respectively.  The difference in the average $P_4/P_0$ between the two samples is more marginal than for $P_3/P_0$ but still fairly significant.  Table 4 summarizes these results and lists the average power ratios for each sample.

\placetable{tbl-4}

Buote \& Tsai (1996) reported a significant correlation between the power ratios of the clusters in their sample.  As can be seen from Figures 2-4, this correlation is also present in our data.  The Spearman Rank-Order Correlation test (e.g., Press et al. 1992, \S 14.6) gives probabilities of 0.017, $9.2\times10^{-6}$, and $2.5\times10^{-5}$ for the $P_2/P_0-P_3/P_0$, $P_2/P_0-P_4/P_0$, and $P_3/P_0-P_4/P_0$ correlations.  A probability of one indicates no correlation, and the test included all 40 clusters.  The same test applied to the low and high-redshift samples separately gave mixed results.  The $P_2/P_0-P_4/P_0$ correlation is significant for both samples, and the $P_3/P_0-P_4/P_0$ correlation is significant for the low-redshift sample.  The correlations among the high-redshift clusters generally gave marginal results, which is perhaps not surprising given the small sample size and relatively large errors.  Similar to Buote \& Tsai (1996), we find the most significant correlation in $P_2/P_0-P_4/P_0$.

Finally, we compare our results to the 59 $z\le0.2$ clusters studied by Buote \& Tsai (1996) with \textit{ROSAT}.  The overall range of power ratios in our sample is very similar to their sample.  A possible problem in comparing these results is the significantly larger PSF of \textit{ROSAT} compared to \textit{Chandra}, which could lead to generally smaller power ratios as measured with \textit{ROSAT}.  However, for the low-redshift clusters studied by Buote \& Tsai (1996) and an aperture radius of 0.5 Mpc, the \textit{ROSAT} PSF should not have large effect.  For the five clusters common to both samples, our power ratios are all contained within the Buote \& Tsai (1996) confidence ranges except $P_4/P_0$ for A1914, and these clusters do not have consistently higher or lower power ratios for \textit{Chandra} versus \textit{ROSAT}.  We find no significant difference between our low-redshift sample and the Buote \& Tsai (1996) sample; however, all three power ratios are significantly higher for our high-redshift sample. Figure 5 shows $P_3/P_0$ versus $P_4/P_0$ for the 59 Buote \& Tsai (1996) clusters compared to our clusters.

\begin{figure}
\centering
\epsscale{0.7}
\plotone{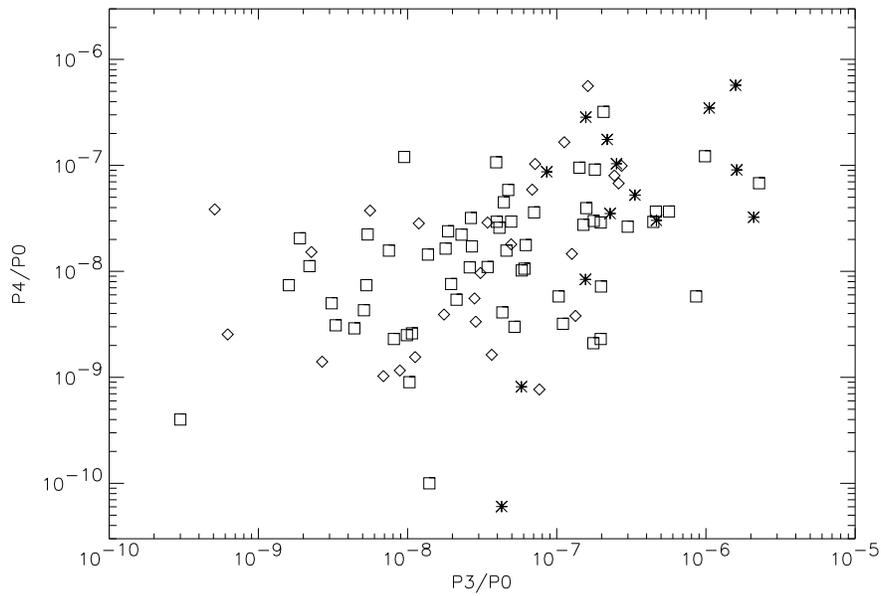}
\figcaption{ $P_3/P_0$ versus $P_4/P_0$ for an aperture radius of 0.5 Mpc.  Plotted with squares are 59 $z\le0.2$ clusters observed with \textit{ROSAT} (Buote \& Tsai 1996).  Diamonds represent our low-redshift sample, and asterisks represent our high-redshift sample. }
\end{figure}

\section{ SYSTEMATICS }

Here we discuss several systematic effects which could influence our results.  

\subsection{ Luminosity and Aperture Radius }

First, we investigate whether the difference in power ratios could be caused by a difference in luminosity between the two samples.  In calculating the power ratios, we normalize by cluster flux to get a ``shape only'' measure; however, one could imagine that massive clusters and groups could have different amounts of substructure.  In selecting the sample, we chose only clusters with relatively high luminosities.  The range of cluster luminosities is only about an order of magnitude, implying that we are comparing reasonably similar clusters, but we will consider further whether the distribution of luminosities is different for the high and low-redshift clusters.  The average luminosities of the two samples are $7.4\times10^{44}$ ergs s$^{-1}$ for the high-redshift sample and $9.1\times10^{44}$ ergs s$^{-1}$ for the low-redshift sample.  This difference in average luminosity is not large and is well within the standard deviations of the two samples.  On the other hand, investigation of Table 1 reveals that a few of the high-redshift clusters have lower luminosities than the rest of the sample.  A KS-test does not show a significant difference between the two samples (probability 0.12), but a rank-sum test gives a probability of 0.047 that the samples have the same mean luminosity.  

We decided to remove the five lowest luminosity clusters from the sample to test whether it would affect our results.  
With these clusters removed, the high-redshift sample has an average luminosity of $10.4\times10^{44}$ ergs s$^{-1}$, and the KS and rank-sum tests of the luminosities give probabilities of 0.54 and 0.42, respectively.  The difference in the average $P_3/P_0$ between the two samples is still very significant, with KS and rank-sum probabilities of 0.013 and 0.0018.  The difference in $P_4/P_0$, on the other hand, is no longer significant at the 0.05 level, giving probabilities for the two tests of 0.17 and 0.10.  While it is still only a marginal result, the high luminosity sample shows a more significant difference in $P_2/P_0$ between the high and low-redshift clusters.  For $P_2/P_0$, the KS-test gives a probability of 0.12 and the rank-sum test gives a probability of 0.051.  $P_3/P_0$ is the most unambiguous indicator of an asymmetric cluster morphology, and the significant difference in $P_3/P_0$ between the two samples even without the low-luminosity clusters shows that the high-redshift clusters tend to have more structure than the low-redshift clusters.  With five clusters removed, the high-redshift sample only contains nine clusters, perhaps leading to the marginal results for $P_4/P_0$.

As a final test of a possible correlation between luminosity and the power ratios, we divided the full cluster sample by luminosity into a low-luminosity half and a high-luminosity half.  After correcting for noise as discussed in the next section, which has a larger effect on the power ratios of the low-luminosity clusters, we find no significant difference in power ratios between the low-luminosity and high-luminosity clusters.  The average $P_3/P_0$ for the low-luminosity sample is a factor of 2.5 greater than for the high-luminosity sample compared to the almost order of magnitude difference in $P_3/P_0$ between the high and low-redshift samples, and the rank-sum and KS-tests give probabilities of 0.40 and 0.77 for a difference in $P_3/P_0$ based on luminosity.  The two luminosity samples show even less of a difference in $P_2/P_0$ and $P_4/P_0$ than they do for $P_3/P_0$.

The second effect we consider is our choice of aperture radius; we chose to use a radius of a fixed physical size rather than a fixed over-density.  A radius of fixed over-density, such as $r_{500}$, would be difficult to determine accurately for both the disturbed and low S/N clusters, but it is worth considering whether we are comparing the same relative scale of structure in the high and low-redshift clusters.  To approximate the difference in physical radius between the high and low-redshift samples that would result from using a radius of fixed over-density we examine our results for an aperture radius of 0.4 Mpc.  For the high-redshift clusters, we substitute the power ratios for $R=0.4$ Mpc and compare these to the power ratios of the low-redshift clusters at $R=0.5$ Mpc.  The shift in the power ratios of high-redshift clusters is generally within the errors and is not consistently either positive or negative.  The 0.4 Mpc and 0.5 Mpc high-redshift power ratios also do not have significantly different means.  Both a rank-sum test and a KS-test show that $P_3/P_0$ is still significantly higher for the high-redshift sample.  A rank-sum test also shows a significant difference in $P_4/P_0$, but the KS probability is 0.10.  We also considered an aperture radius of 0.3 Mpc for the high-redshift clusters compared to a radius of 0.5 Mpc for the low-redshift clusters, because this radius is closer to the correct comparison radius for the $z>0.8$ clusters.  We found that all three power ratios were then significantly higher for the high-redshift sample.

\subsection{ Noise }

Finally, we look into the effect of noise, because the high-redshift clusters generally have lower S/N than the low-redshift clusters.  As a test, we first experimented with both lowering the S/N of a few of our cluster observations and with adding noise to model clusters with a range of morphologies.  We find that for clusters with distinct substructure and relatively large power ratios, noise has little effect as it only contributes a few percent to the power ratios.  For fairly relaxed clusters with small power ratios, the effects of noise are more important.  Noise can artificially inflate the power ratios in this case.

We then employed an analytical method to calculate the expected contribution of noise to the power ratios for each cluster.  The contribution of noise to the square of the moments is taken to be
\begin{eqnarray}
a_{m,noise}^2(R) & = & \int_{R^{\prime}\le R} \Sigma(\vec x^{\prime})\left(R^{\prime}\right)^{2m} \cos^2 m\phi^{\prime} d^2x^{\prime} + \int_{R^{\prime}\le R} {\Sigma_b(\vec x^{\prime})\over A}\left(R^{\prime}\right)^{2m} \cos^2 m\phi^{\prime} d^2x^{\prime}, \\
b_{m,noise}^2(R) & = & \int_{R^{\prime}\le R} \Sigma(\vec x^{\prime})\left(R^{\prime}\right)^{2m} \sin^2 m\phi^{\prime} d^2x^{\prime} + \int_{R^{\prime}\le R} {\Sigma_b(\vec x^{\prime})\over A}\left(R^{\prime}\right)^{2m} \sin^2 m\phi^{\prime} d^2x^{\prime}, 
\end{eqnarray}
where $\Sigma$ is the observed surface brightness, $\Sigma_b$ is the surface brightness due to the background, and $A$ is the exposure normalization applied to the background images.  The derivation of these equations can be found in Appendix B.  We applied these formulas to the adaptively binned images of each cluster to estimate the net increase in the power ratios due to noise and then subtracted these noise contributions from the observed ratios.  The noise corrected power ratios are listed in Table 5.  After the noise correction, some of the net power ratios are negative, which is unphysical but valid for our comparison of cluster ratios.  

The difference in $P_3/P_0$ between the high and low-redshift samples remains significant with rank-sum and KS-test probabilities of 0.013 and 0.015, respectively.  However, we no longer find a significant difference in $P_4/P_0$ between the two samples.  Given our somewhat small sample size, particularly at high-redshift, we employed a bootstrap resampling method to test the effects of sample size and selection.  The clusters were randomly resampled with replacement, meaning that some clusters will be selected multiple times and other will not be selected at all, for both the high and low-redshift samples, and the rank-sum and KS-tests for $P_3/P_0$ were repeated.  For 1000 resamplings more than 70\% of the resamplings gave probabilities less than 0.05 for both tests with average probabilities for the two tests of 0.062 and 0.053.  Our results are therefore fairly robust to sample selection and our sample size is reasonable, but as expected it does have some effect.  This situation will improve as more high-redshift clusters are observed.  The average power ratios, rank-sum, and KS-test probabilities after noise correction are listed in Table 6.  Figure 6 shows the three projections of the power ratios after both the ratios and their errors have been corrected for noise.  Where the ratios are negative, they have been plotted at their upper limits with an arrow indicating the limit.

\placetable{tbl-5}
\placetable{tbl-6}

Given the observed significant difference in $P_3/P_0$ between the high and low-redshift samples, we sought to fit the slope of this evolution in structure.  Using a least absolute deviation method (e.g., Press et al. 1992, \S 15.7), we fit a line to redshift versus noise-corrected $P_3/P_0$.  To incorporate the uncertainties in the power ratios, we randomly selected values of $P_3/P_0$ for each cluster from the Monte Carlo simulations, as in section 5, and repeated the fit 1000 times.  We find an average slope of $4.09\times10^{-7}$ and 90\% confidence limits on the slope of $8.19\times10^{-8}-8.03\times10^{-7}$, where the limits were taken to be the fiftieth lowest and fiftieth highest slopes.  Only 5 of the 1000 fits gave negative slopes, confirming the significant evolution in $P_3/P_0$ with redshift.  A least-squares fitting method gave even larger values of the slope.  We also fit redshift versus noise-corrected $P_4/P_0$ using the same procedure.  For the least absolute deviation method, we found an average slope of $7.47\times10^{-8}$; however, 161 of 1000 fits gave negative slopes.  The average slopes and their confidence limits are also listed in Table 6.

Longer observations of the high-redshift, single clusters would help to better determine their structure and place stronger constraints on structure evolution. 
Buote \& Tsai (1995) also looked at the effect of noise on the power ratios of model clusters.  Their results show that for cluster models with small or zero power ratios the addition of noise significantly increases the power ratios, but not enough to make a relaxed cluster look like a disturbed cluster.  For a cluster model with a perfectly round surface brightness distribution the increase in the power ratios after the addition of noise above the true value of zero is roughly consistent with our estimates of the effect of noise on our observed clusters for clusters with a similar signal-to-noise.

\begin{figure}
\centering
\epsscale{0.5}
\plotone{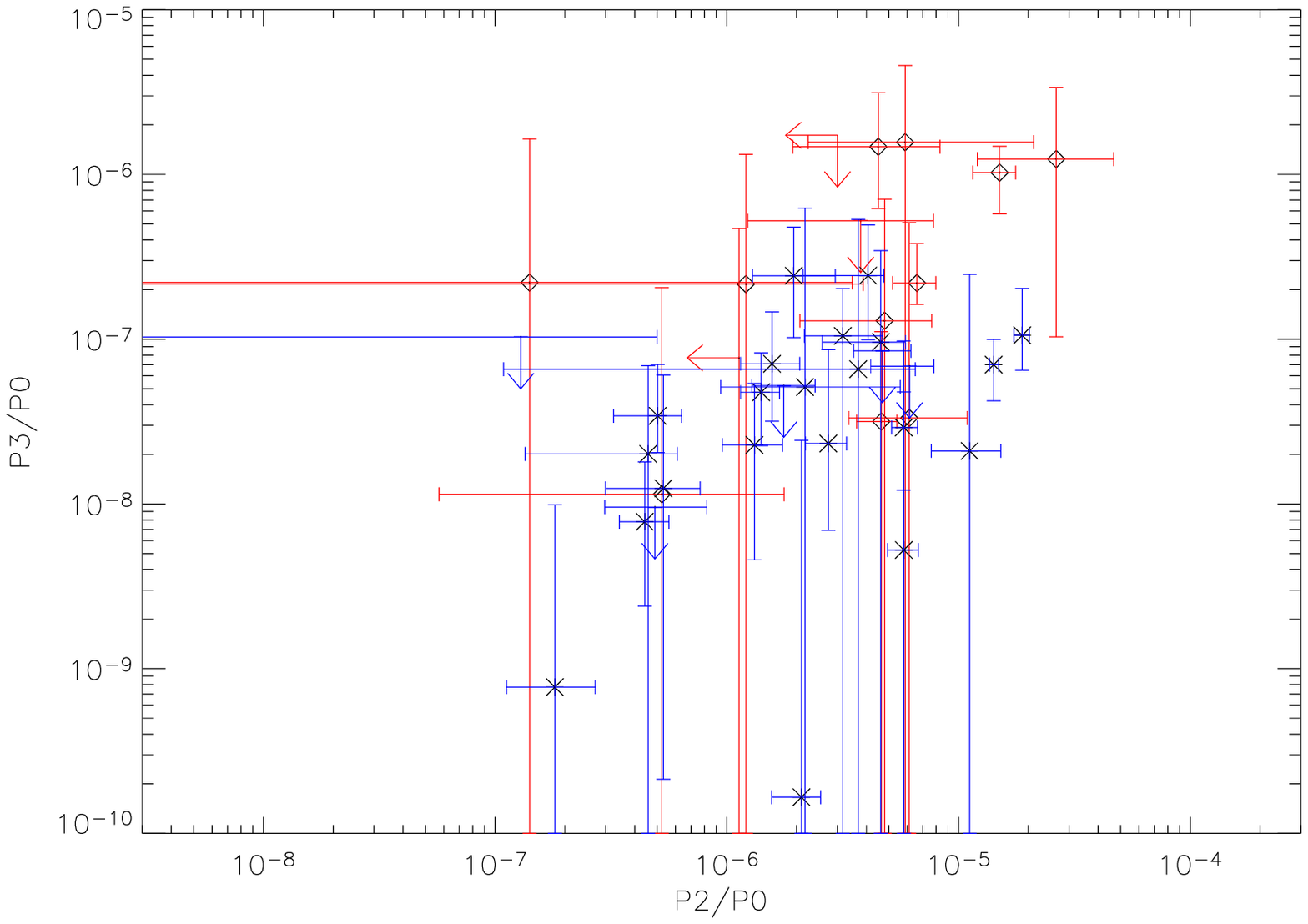}
\plotone{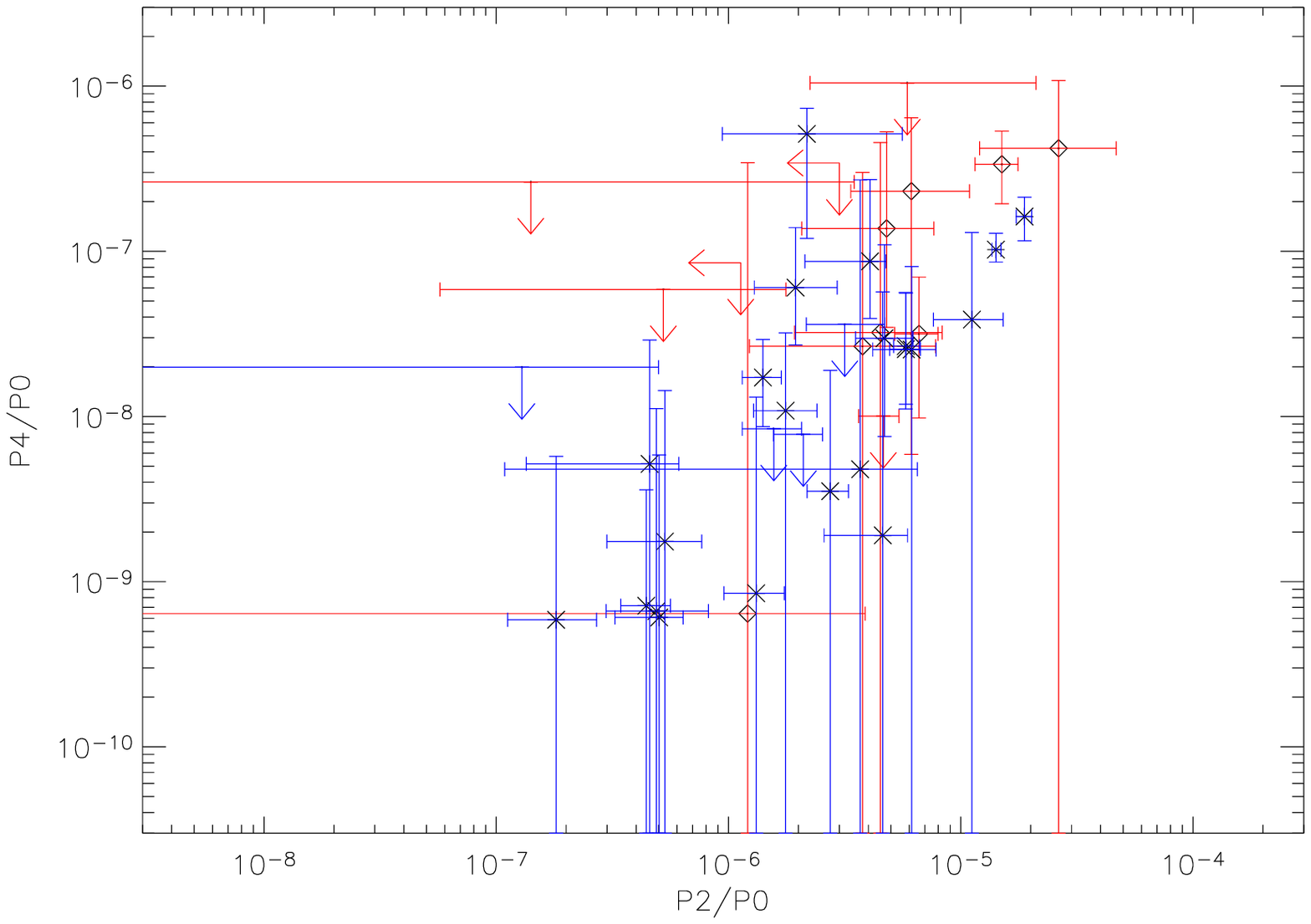}
\plotone{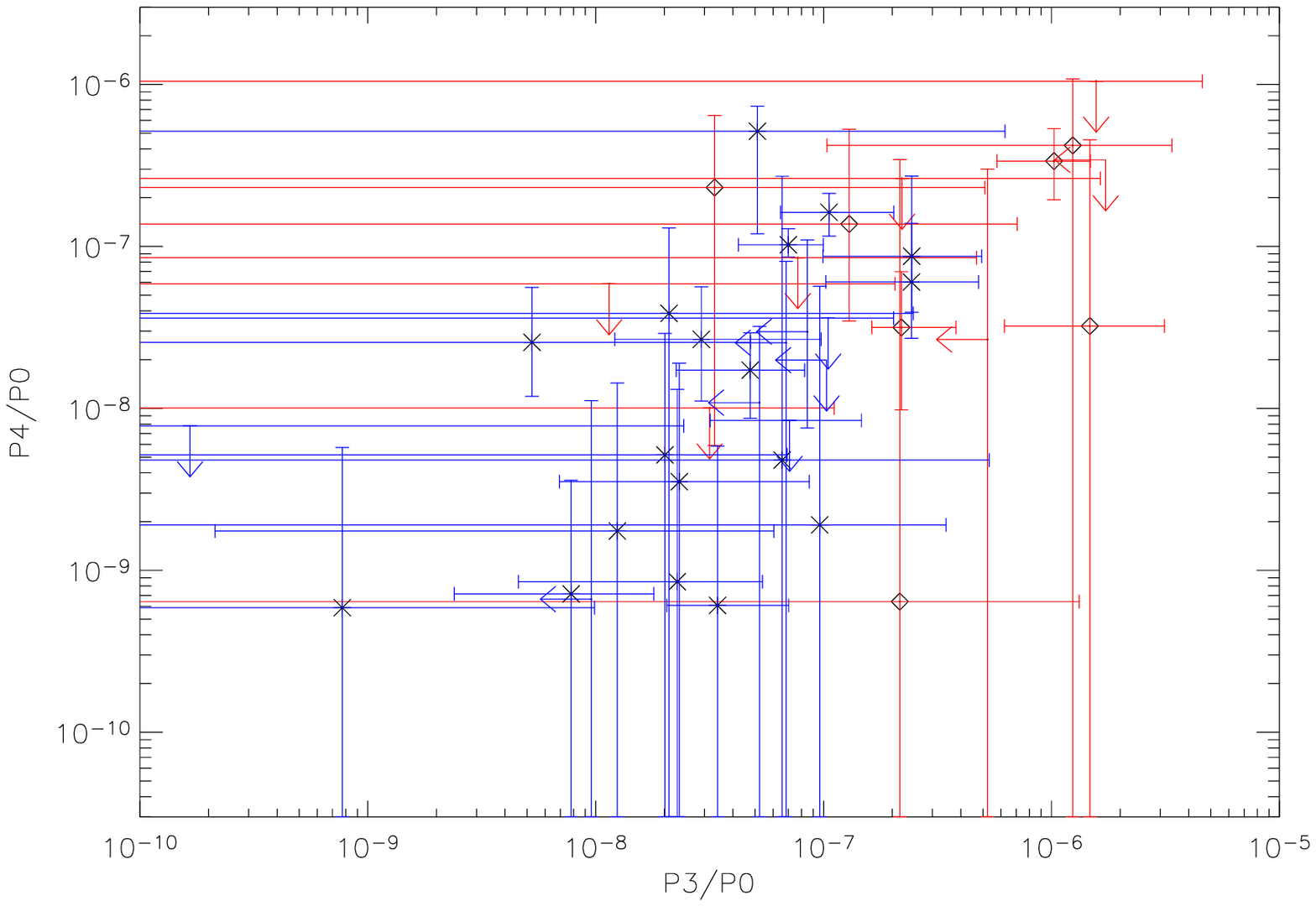}
\figcaption{ Noise corrected power ratios and uncertainties.  Where the power ratios are negative, they have been plotted with arrows at their upper limits.  Top: $P_2/P_0$ versus $P_3/P_0$.  Middle: $P_2/P_0$ versus $P_4/P_0$.  Bottom: $P_3/P_0$ versus $P_4/P_0$.  High-redshift clusters ($z>0.5$) are plotted with diamonds and have red error bars.  Low-redshift clusters are plotted with asterisks and have blue error bars. }
\end{figure}

To summarize, we have tested three possible sources of systematic error: the inclusion of a few lower luminosity high-redshift clusters in our sample, our use of an aperture radius of fixed physical size versus fixed over-density, and the generally lower S/N of the high-redshift cluster data compared to the low-redshift data.  The fact that the high-redshift clusters typically have higher $P_3/P_0$ than the low-redshift clusters remains highly significant under all of these tests.  The difference in $P_4/P_0$ between the two samples is still significant for the change in aperture radius but is marginal for the other two tests.  We also fit redshift versus the noise-corrected values of $P_3/P_0$ and find a significant positive slope.

\section{ DISCUSSION AND CONCLUSIONS }

We investigate the observed evolution of cluster structure with redshift out to $z\sim1$ using a sample of 40 clusters observed with \textit{Chandra}.  We find that, as expected from hierarchical models of structure formation, high-redshift clusters have more substructure and are dynamically younger than low-redshift clusters.  Specifically, the clusters in our sample with $z>0.5$ tend to have both higher $P_3/P_0$ and higher $P_4/P_0$ than the clusters with $z<0.5$.  We do not find a significant difference in $P_2/P_0$ between the two samples.  The results for $P_3/P_0$ are robust when compared to the effects of noise, luminosity, and the choice of aperture radius.  The results for $P_4/P_0$ are also fairly robust but show more sensitivity to these effects.  As even multipoles, both $P_2/P_0$ and $P_4/P_0$ are sensitive to ellipticity.  Ellipticity is not a clear indicator of dynamical state; a double cluster and a relaxed, but elliptical single cluster can have the same ellipticity.  A high $P_3/P_0$ is an unambiguous indicator of an asymmetric cluster structure, and the evolution of $P_3/P_0$ remains highly significant despite all of the possible systematic effects that we considered.  Given the observation of a significant positive evolution, we fit for the slope of this effect.  Using a least absolute deviation fitting method and randomly selecting values of $P_3/P_0$ for each cluster from Monte Carlo simulations, we fit redshift versus $P_3/P_0$, after noise correction, and find an average slope of $4.09\times10^{-7}$ and 90\% confidence limits of $8.19\times10^{-8}-8.03\times10^{-7}$.  The slope is greater than zero at better than 99\% confidence.

The observation of structure evolution in the redshift range probed by current cluster surveys indicates that dynamical state should be taken into account by cosmological cluster studies. Mergers can lead to large deviations in cluster luminosity, temperature, and velocity dispersion (Rowley, Thomas, \& Kay 2004; Randall, Sarazin, \& Ricker 2002; Mathiesen \& Evrard 2001) and therefore errors in estimating cluster mass and gas mass fraction as well as possible selection effects in cluster surveys.  In the future, we will use these results to place constraints on cosmological models by comparing the observed evolution in cluster morphology to the evolution predicted by hydrodynamic simulations.  Alternatively, this comparison can be seen as a test of whether current simulations reproduce the observed structure in clusters on the scales probed by the power ratios.  Our findings and method may also prove useful in understanding the relationship of dynamical state to other observed cluster properties such as galaxy evolution and strong lensing.  Increased star-formation rates have been observed in clusters undergoing mergers (e.g., Miller \& Owen 2003; Metevier, Romer, \& Ulmer 2000; Caldwell \& Rose 1997), and an increase in merging with redshift could contribute to the Butcher-Oemler effect.  Gladders et al. (2003) found a high incidence of giant arcs at high-redshift in the Red Sequence Cluster Survey, which could also be influenced by cluster structure.

\acknowledgements
We would like to thank E. Bertschinger, S. Burles, and H. Marshall for useful discussions, as well as the MIT CXC group.  This work was supported by NASA through contract NAS8-01129.

\appendix
\section{ NOTES ON INDIVIDUAL CLUSTERS }

Here we give a brief description of the clusters in our sample.  Also listed are any specific notes on their processing.  For the clusters that were not selected from either the EMSS or BCS, we indicate the survey in which they were discovered.  Images of all of the clusters can be found in Appendix C.

\noindent \textbf{MS0015.9+1609} ($z=0.54$) - This cluster is elliptical and has an asymmetric core structure. 

\noindent \textbf{CLJ0152.7-1357} ($z=0.83$) - This cluster is a well-separated double cluster with nearly equal mass subclusters.  Discovered in a number of surveys including RDCS (Rosati et al. 1998), WARPS (Perlman et al. 2002), and the bright SHARC (Romer et al. 2000).  It was also discovered by \textit{Einstein}, but its significance was underestimated due to its morphology, and it was not included in the EMSS cluster catalog.

\noindent \textbf{A267} ($z=0.23$) - A267 appears to be a disturbed, single component cluster.  Its core is very elliptical, offset from the center, and twists from being extended in the NE-SW direction to extending S.  The cluster as a whole is also extended in the NE-SW direction.  A smaller extended source appears 2 Mpc east of the cluster.  There were three observations of this cluster in the archive, but one was very short and another had a high background count rate.  We used only the longest observation.  

\noindent \textbf{MS0302.7+1658} ($z=0.42$) - This cluster is elliptical, the core is offset to the south, and it has excess emission to the NW making the outer contours look triangular.

\noindent \textbf{RXJ0439.0+0715} ($z=0.24$) - RXJ0439.0+0715 is a single, elliptical cluster.  The core is offset a bit from the center of the cluster, but otherwise it appears fairly symmetric.  This cluster had three observations in the archive.  However, one of these observations was for less than a kilosecond, and another observation had a high background rate.  We only used the longest of the three observations.

\noindent \textbf{RXJ0439.0+0520} ($z=0.21$) - A fairly round, relaxed looking single cluster.  For this cluster, the 0.5 Mpc aperture fell on two of the ACIS-I CCDs.

\noindent \textbf{A520} ($z=0.20$) - This cluster has clear substructure in the form of a bright arm of emission leading to a bright knot to the SW.  This clump is cooler than the surrounding gas, and it appears to be a group of galaxies that has recently passed through the main cluster (Govoni et al. 2004).

\noindent \textbf{MS0451.6-0305} ($z=0.54$) - This cluster is elliptical and extended in a roughly E-W direction, with excess emission to the E.  The core is lumpy and extended, and there is evidence that the BCG has not yet settled into the center of the cluster (Donahue et al. 2003).  There were two observations of this cluster in the archive, but both the instrument and the focal plane temperature differed between the observations.  We chose to use only the longer, ACIS-S observation.

\noindent \textbf{CLJ0542.8-4100} ($z=0.63$) - This cluster is extended in the E-W direction.  The core is elliptical, offset from the center, and shows a position angle twist from E-W to SE.

\noindent \textbf{A665} ($z=0.18$) - A665 is a very odd looking cluster with a twisted, almost S-shaped core and a large extension to the NW.  Several observation suggest that this cluster is currently undergoing a merger (Govoni et al. 2004; Gomez, Hughes, \& Birkinshaw 2000; Buote \& Tsai 1996).  The 0.5 Mpc aperture fell on two of the ACIS-I CCDs.

\noindent \textbf{ZWCL1953} ($z=0.38$) - This cluster has a double peaked core which is offset from the centroid.  It is also extended in a roughly N-S direction.

\noindent \textbf{MS0906.5+1110} ($z=0.18$) - This cluster is elliptical and fairly relaxed with some possible extension in the core.  There is a second extended structure at the edge of the CCD which may or may not be associated with MS0906.5+1110.  However, this component does not fall in the 0.5 Mpc aperture.

\noindent \textbf{A773} ($z=0.22$) - A773 is roughly elliptical and has a very elongated, slightly curved core.  The temperature map and galaxy distribution suggest that it is undergoing a merger (Govoni et al. 2004).  In this observation, the 0.5 Mpc aperture fell on two of the ACIS-I CCDs.

\noindent \textbf{A781} ($z=0.30$) - This is a complex cluster with multiple peaks.  The core is offset from the center and has an extension that curves toward a peak to the north.  There is also an extension toward a peak to the east, but the peak itself falls outside the 0.5 Mpc aperture.  To the E-SE there are two other extended X-ray sources which may form part of line of clusters along a filament.  The closest of these is over 1.2 Mpc from the center of A781.

\noindent \textbf{A963} ($z=0.21$) - A963 is elliptical and very relaxed looking with some possible very small scale structure in the core.  We included the effects of background noise when calculating the uncertainties in the power ratios of this cluster.

\noindent \textbf{ZWCL3146} ($z=0.29$) - This cluster is elliptical and fairly relaxed and has a slightly offset core.  ZWCL3146 is a cooling flow cluster with a complicated core structure (Forman et al. 2003), but this structure is very small scale and therefore not important to this analysis.  The 0.5 Mpc aperture fell on two of the ACIS-I CCDs.

\noindent \textbf{MS1054.5-0321} ($z=0.83$) - MS1054.5-0321 is a nearly equal mass, double cluster and the highest redshift cluster in the EMSS.

\noindent \textbf{CLJ1113.1-2615} ($z=0.73$) - A single, relatively relaxed looking cluster.  
Discovered in WARPS.  For this cluster, the background moments were similar to the cluster moments, and we included the effects of background noise in the calculation of the uncertainties.

\noindent \textbf{V1121.0+2327} ($z=0.56$) - V1121.0+2327 is a complicated cluster with 2-3 peaks and very twisted emission.  Discovered in the 160 deg$^2$ survey (Vikhlinin et al. 1998).  This cluster has background moments similar to the cluster moments, and we included the effects of background noise in the uncertainties.

\noindent \textbf{MS1137.5+6625} ($z=0.78$) - A relaxed looking single cluster.  We included the effects of background noise in the calculation of the uncertainties for this cluster.  

\noindent \textbf{A1413} ($z=0.14$) - An elliptical, but otherwise relaxed looking cluster.  There were two observations of this cluster in the archive, but one of these observations had a high background rate.

\noindent \textbf{V1221.4+4918} ($z=0.70$) - This cluster has two peaks surrounded by an outer envelope that is extended in the NW-SE direction.  Discovered in the 160 deg$^2$ survey.

\noindent \textbf{CLJ1226.9+3332} ($z=0.89$) - CLJ1226.9+3332 is the highest redshift cluster in our sample, and it is a fairly relaxed single cluster.  
Maughan et al. (2004) observed this cluster with \textit{XMM-Newton}.  They find it to be very hot and fairly isothermal.  There are possibly two other extended sources near this cluster, one 1.8 Mpc to the east and one 4.5 Mpc to the north.  CLJ1226.9+3332 was discovered in WARPS.  We included the effects of background noise in the calculation of the uncertainties in the power ratios for this cluster.

\noindent \textbf{A1758} ($z=0.28$) - One of the most complicated looking low-redshift clusters.  A1758 has approximately two main clumps but these are lumpy and disrupted, probably due to an ongoing merger where the two clusters have already passed through each other.  The outer envelope of this cluster is elliptical.

\noindent \textbf{RXJ1350.0+6007} ($z=0.80$) - A disturbed looking single cluster with an offset core and a large position angle twist from E-W to SE.  Discovered in the RDCS.  For this cluster, the background moments were similar to the cluster moments, and we included the effects of background noise in the calculation of the uncertainties.

\noindent \textbf{MS1358.4+6245} ($z=0.33$) - MS1358.4+6245 is a relaxed, cooling flow cluster (Arabadjis, Bautz, \& Garmire 2002).  In the smoothed image, it is elliptical and has some extension in the core; however, this core structure is fairly small scale.  We included the effects of background noise in the calculation of the uncertainties.

\noindent \textbf{A1914} ($z=0.17$) - This cluster has two peaks surrounded by an outer roughly elliptical envelope.  The SE peak extends and curves toward the NW peak.  The temperature map shows a hot region between the two peaks suggesting shock heated gas (Govoni et al. 2004).  In this observation, the 0.5 Mpc aperture fell on two of the ACIS-I CCDs.

\noindent \textbf{A2034} ($z=0.11$) - A2034 is fairly round but has some small scale core structure, including a curved structure extending away from the peak.  It also shows both a northern cold front and a southern excess (Kempner, Sarazin, \& Markevitch 2003).  It is currently unclear if the southern excess is associated with A2034, and Kempner et al. (2003) suggest that it may be a background cluster.  Most of the southern excess is not in the 0.5 Mpc aperture, but our aperture does overlap the top of this feature.  Comparing to the 0.3 Mpc and 0.4 Mpc apertures which do not contain the excess, $P_3/P_0$ and $P_4/P_0$ do shift somewhat.  However, this shift is toward smaller values of the power ratios which would only strengthen our conclusions.  This cluster has the lowest redshift in our sample, and the 0.5 Mpc aperture overlapped all four ACIS-I CCDs.

\noindent \textbf{RXJ1532.9+3021} ($z=0.35$) - Relaxed, slightly elliptical single cluster.  There is some very small scale extension in the core of this cluster.  There were two 10 ksec observations of RXJ1532.9+3021, one with ACIS-I and one with ACIS-S.  We used only the ACIS-S observation which appeared to have a higher net cluster count rate.

\noindent \textbf{A2111} ($z=0.23$) - This cluster is very elongated and extended toward the NW.  It also has a double peaked core.

\noindent \textbf{A2218} ($z=0.18$) - A2218 looks fairly relaxed and symmetric except for a slightly offset core and some very small scale core structure.  Govoni et al. (2004) find an asymmetric temperature structure, and they suggest that this cluster is a late-stage merger.  This cluster had three observations in the archive, including two short observations at a focal plane temperature of $-110^{\circ}$C and a longer observation at $-120^{\circ}$C.  We used only the last observation which had more than double the exposure time of the other two.  The 0.5 Mpc aperture fell on three of the ACIS-I CCDs.

\noindent \textbf{A2219} ($z=0.23$) - This cluster is very elliptical and elongated in the NW-SE direction.  It also has some small scale lumpiness in the center.  This cluster has both an elongated galaxy distribution and an elongated radio halo (Boschin et al. 2004).  Boschin et al. (2004) suggest that A2219 is a late-stage merger.

\noindent \textbf{RXJ1716.9+6708} ($z=0.81$) - This cluster has a small subcluster or group to the NE of the main cluster.  In addition, the core is elongated in the direction of the subcluster.  In the NASA/IPAC Extragalactic Database (NED), we found two cluster galaxies associated with this small clump.  RXJ1716.9+6708 was discovered in the NEP (Gioia et al. 2003).  The optical galaxy distribution resembles an inverted S-shaped filament (Gioia et al. 1999).

\noindent \textbf{RXJ1720.1+2638} ($z=0.16$) - A fairly round, relaxed cluster.  It becomes somewhat elliptical outside the 0.5 Mpc aperture.  There were three observations of this cluster, one was very short and the other two were at different focal plane temperatures.  We use only the longest observation performed at a focal plane temperature of $-110^{\circ}$C.  The 0.5 Mpc aperture fell on three of the ACIS-I CCDs.

\noindent \textbf{A2261} ($z=0.22$) - This cluster has a small secondary clump to the west of the main cluster.  This clump does not fall in the 0.5 Mpc aperture, and within the aperture A2261 is very round and relaxed.

\noindent \textbf{MS2053.7-0449} ($z=0.58$) - A single cluster; it is extended in the NW-SE direction. 
For this cluster, we merged two observations.  Both observations were made with ACIS-I at a focal plane temperature of $-120^{\circ}$C, one in F mode and one in VF mode.  The two observations were aligned by hand using six bright X-ray point sources that appear in both images.

\noindent \textbf{RXJ2129.6+0005} ($z=0.24$) - An elliptical, relaxed looking single cluster.  It is elongated in the NE-SW direction.

\noindent \textbf{MS2137.3-2353} ($z=0.31$) - MS2137.3-2353 is a very relaxed looking, slightly elliptical cluster.

\noindent \textbf{A2390} ($z=0.23$) - This cluster has a large scale extension to the east as well as a pointy or triangular extension to the NW, giving it an odd appearance.  It also has some very small scale structure in the core.  A2390 was observed twice with \textit{Chandra}, but we use only the observation at a focal plane temperature of $-120^{\circ}$C.

\noindent \textbf{CLJ2302.8+0844} ($z=0.73$) - This cluster is a fairly relaxed looking single cluster.  There is a large foreground galaxy near this cluster, but it is not in the 0.5 Mpc aperture.  Discovered in WARPS.  For this cluster, the background moments were similar to the cluster moments, and we included the effects of background noise in the calculation of the uncertainties.

\section{CALCULATION OF THE NOISE CONTRIBUTION TO THE POWER RATIOS}

Let $o_i$ be the surface brightness in pixel $i$ of the cluster image, where $o_i = c_i + b_i$ is the sum of the observed cluster and background surface brightnesses, and let $\beta_i$ be the surface brightness in pixel $i$ in the background image before normalization.  We assume that $c_i$, $b_i$, and $\beta_i$ are Poisson random variables with means equal to their true values.  The expected value of $o_i$ is then
\begin{equation}
\langle o_i \rangle=C_i+B_i,
\end{equation}
where $C_i$ is the true cluster surface brightness in pixel $i$, and $B_i$ is the true background surface brightness in pixel $i$.  In addition,
\begin{equation}
\langle \beta_i \rangle=B_iA,
\end{equation}
where $A$ is the normalization for exposure between the cluster observation and the background image.  The expected values of the moments then are
\begin{eqnarray}
\langle a_m \rangle & = & \sum_{i=1}^{n} \langle o_i \rangle \left(R^{\prime}_i\right)^m \cos m\phi_i - \sum_{i=1}^{n} {\langle \beta_i \rangle \over A}\left(R^{\prime}_i\right)^m \cos m\phi_i \nonumber \\
    & = & \sum_{i=1}^{n} C_i\left(R^{\prime}_i\right)^m \cos m\phi_i \\
\langle b_m \rangle & = & \sum_{i=1}^{n} \langle o_i \rangle \left(R^{\prime}_i\right)^m \sin m\phi_i - \sum_{i=1}^{n} {\langle \beta_i \rangle \over A}\left(R^{\prime}_i\right)^m \sin m\phi_i \nonumber \\
    & = & \sum_{i=1}^{n} C_i\left(R^{\prime}_i\right)^m \sin m\phi_i,
\end{eqnarray}
where $\sum_{i=1}^{n}$ indicates a sum over all pixels in a circular aperture with radius $R$.  It follows that the expected value of $a_m^2$ is
\begin{eqnarray}
\langle a_m^2 \rangle & = & \sum_{i,j=1}^{n} \left(\langle c_ic_j \rangle + \langle c_ib_j \rangle - {\langle c_i\beta_j \rangle \over A} + \langle b_ic_j \rangle + \langle b_ib_j \rangle - {\langle b_i\beta_j \rangle \over A} \right.\nonumber \\
      &  & \left. \mbox{} - {\langle \beta_ic_j \rangle \over A} - {\langle \beta_ib_j \rangle \over A} + {\langle \beta_i\beta_j \rangle \over A^2} \right)\left(R^{\prime}_i\right)^m\left(R^{\prime}_j\right)^m \cos m\phi_i \cos m\phi_j.
\end{eqnarray}
Assuming that different pixels are independent (i.e. $\langle c_ic_j \rangle = \langle c_i \rangle \langle c_j \rangle$) and using the property \\$\langle c_i^2 \rangle =  \langle c_i(c_i - 1) \rangle + \langle c_i \rangle = C_i^2 +C_i$ for Poisson random variables
\begin{eqnarray}
\sum_{i,j=1}^{n} \langle c_ic_j \rangle\left(R^{\prime}_i\right)^m\left(R^{\prime}_j\right)^m \cos m\phi_i \cos m\phi_j & = & \sum_{i,j=1}^{n} C_iC_j\left(R^{\prime}_i\right)^m\left(R^{\prime}_j\right)^m \cos m\phi_i \cos m\phi_j \nonumber \\
       &  & \mbox{} + \sum_{i=1}^{n} C_i\left(R^{\prime}_i\right)^{2m} \cos^2 m\phi_i \nonumber \\
\sum_{i,j=1}^{n} \langle b_ib_j \rangle\left(R^{\prime}_i\right)^m\left(R^{\prime}_j\right)^m \cos m\phi_i \cos m\phi_j & = & \sum_{i,j=1}^{n} B_iB_j\left(R^{\prime}_i\right)^m\left(R^{\prime}_j\right)^m \cos m\phi_i \cos m\phi_j \nonumber \\
       &  & \mbox{} + \sum_{i=1}^{n} B_i\left(R^{\prime}_i\right)^{2m} \cos^2 m\phi_i \nonumber \\
\sum_{i,j=1}^{n} {\langle \beta_i\beta_j \rangle \over A^2}\left(R^{\prime}_i\right)^m\left(R^{\prime}_j\right)^m \cos m\phi_i \cos m\phi_j & = & \sum_{i,j=1}^{n} B_iB_j\left(R^{\prime}_i\right)^m\left(R^{\prime}_j\right)^m \cos m\phi_i \cos m\phi_j \nonumber \\
       &  & \mbox{} + \sum_{i=1}^{n} {B_i \over A}\left(R^{\prime}_i\right)^{2m} \cos^2 m\phi_i \nonumber
\end{eqnarray}
\begin{eqnarray}
\langle c_ib_j \rangle & = & {\langle c_i\beta_j \rangle \over A} \nonumber \\
\langle b_ic_j \rangle & = & {\langle \beta_ic_j \rangle \over A} \nonumber \\
{\langle b_i\beta_j \rangle \over A} & = & {\langle \beta_ib_j \rangle \over A} = B_iB_j. \nonumber
\end{eqnarray}
Substituting the above into equation B5, the expected value of $a_m^2$ reduces to
\begin{equation}
\langle a_m^2 \rangle = \sum_{i,j=1}^{n} C_iC_j\left(R^{\prime}_i\right)^m\left(R^{\prime}_j\right)^m \cos m\phi_i \cos m\phi_j + \sum_{i=1}^{n} \left(C_i + B_i + {B_i \over A}\right)\left(R^{\prime}_i\right)^{2m} \cos^2 m\phi_i.
\end{equation}
The first term in this equation is simply the true $a_m^2$ of the cluster while the last three terms are the additional contribution to the power from noise.  Estimating the true surface brightnesses as the observed surface brightnesses, $\Sigma_i = C_i +B_i$ and $\Sigma_{b,i} = B_i$, and transforming to an integral, we recover equation 5 for the noise contribution to $a_m^2$.  Equation 6 for $b_{m,noise}^2$ is derived similarly.

\section{ CLUSTER IMAGES }

In this appendix, we show smoothed \textit{Chandra} images of the clusters in our sample in order of increasing redshift.  These images were created using the CIAO program \textit{csmooth}, and they are approximately 1.4 Mpc on a side.  The point source removal and exposure correction have not been applied to these images.  For those clusters where the aperture used to calculate the power ratios fell on multiple CCDs the low exposure regions (chip gaps and some bad columns) of the image have been masked out.

\noindent(For a version of the paper including cluster images see http://www.ociw.edu/$\sim$tesla/structure.ps.gz)

\begin{figure}
\epsscale{0.85}
\end{figure}

\begin{figure}
\epsscale{0.85}
\end{figure}

\begin{figure}
\epsscale{0.85}
\end{figure}

\begin{figure}
\epsscale{0.85}
\end{figure}

\begin{figure}
\epsscale{0.85}
\end{figure}

\begin{figure}
\epsscale{0.85}
\end{figure}

\begin{figure}
\epsscale{0.85}
\end{figure}

\begin{deluxetable}{lcccc}
\tablecaption{ The Sample }
\tablewidth{0pt}
\tablecolumns{5}
\tablehead{
\colhead{ Cluster } & \colhead{ Redshift } & 
\colhead{ObsID} & \colhead{ Exposure (ks) } &
\colhead{ Luminosity } \\
& & & & \colhead{($10^{44}h_{70}^{-2}$ ergs s$^{-1}$)}
}
\startdata
MS0015.9+1609 &0.54 &520 &66 &16. \\
CLJ0152.7-1357 &0.83 &913 &36 & 5.0 \\
A267 &0.23 &1448 &7.0 &6.0 \\
MS0302.7+1658 &0.42 &525 &9.9 &2.8 \\
RXJ0439.0+0715 &0.24 &1449 &6.2 &7.7 \\
RXJ0439.0+0520 &0.21 &527 &9.5 &4.6 \\
A520 &0.20 &528 &9.3 &6.2 \\
MS0451.6-0305 &0.54 &902 &42 &17. \\
CLJ0542.8-4100 &0.63 &914 &50 &4.2 \\
A665 &0.18 &531 &8.9 &7.6 \\
ZWCL1953 &0.38 &1659 &18 &9.0 \\
MS0906.5+1110 &0.18 &924 &29 &3.9 \\
A773 &0.22 &533 &11 &7.2 \\
A781 &0.30 &534 &9.6 &4.4 \\
A963 &0.21 &903 &36 &7.9 \\
ZWCL3146 &0.29 &909 &45 &23. \\
MS1054.5-0321 &0.83 &512 &75 &14. \\
CLJ1113.1-2615 &0.73 &915 &96 &2.0 \\
V1121.0+2327 &0.56 &1660 &69 &2.4 \\
MS1137.5+6625 &0.78 &536 &115 &6.6 \\
A1413 &0.14 &1661 &8.8 &7.5 \\
V1221.4+4918 &0.70 &1662 &76 &4.8 \\
CLJ1226.9+3332 &0.89 &932 &9.2 &20. \\
A1758 &0.28 &2213 &48 &7.9 \\
RXJ1350.0+6007 &0.80 &2229 &57 &2.3 \\
MS1358.4+6245 &0.33 &516 &47 &7.1 \\
A1914 &0.17 &514 &6.9 &14. \\
A2034 &0.11 &2204 &52 &4.0 \\
RXJ1532.9+3021 &0.35 &1649 &9.2 &20. \\
A2111 &0.23 &544 &10 &4.4 \\
A2218 &0.18 &1666 &34 &5.4 \\
A2219 &0.23 &896 &42 &16. \\
RXJ1716.9+6708 &0.81 &548 &51 &5.6 \\
RXJ1720.1+2638 &0.16 &1453 &7.7 &9.9 \\
A2261 &0.22 &550 &8.4 &11. \\
MS2053.7-0449 &0.58 &1667, 551 &44, 44 &2.2 \\
RXJ2129.6+0005 &0.24 &552 &9.8 &9.7 \\
MS2137.3-2353 &0.31 &928 &31 &14. \\
A2390 &0.23 &500 &9.7 &17. \\
CLJ2302.8+0844 &0.73 &918 &105 &2.1 \\
\enddata
\tablecomments{ Column 4 gives the net exposure after the removal of background flares.  Column 5 lists cluster luminosity in the 0.3-7.0 keV band estimated using a Raymond-Smith thermal plasma model with $N_H=3\times10^{20}$ atoms cm$^{-2}$, $kT=5$ keV, and an abundance of 0.3. Luminosities are calculated in a circular aperture with a radius of 0.5 Mpc.}
\end{deluxetable}

\begin{deluxetable}{lccc}
\tablecaption{ Power Ratios }
\tablewidth{0pt}
\tablecolumns{4}
\tablehead{
\colhead{ Cluster } & \colhead{$P_2/P_0$ ($\times10^{-7}$)} & 
\colhead{$P_3/P_0$ ($\times10^{-7}$)} & \colhead{$P_4/P_0$ ($\times10^{-7}$)}
}
\startdata
MS0015.9+1609 &46.8 &0.0427 &0.000603 \\
CLJ0152.7-1357 &275. &15.8 &5.70 \\
A267 &62.5 &0.00510 &0.384 \\
MS0302.7+1658 &42.8 &2.44 &0.801 \\
RXJ0439.0+0715 &47.2 &1.26 &0.146 \\
RXJ0439.0+0520 &1.82 &0.112 &0.0155 \\
A520 &41.7 &2.72 &0.991 \\
MS0451.6-0305 &66.4 &2.28 &0.351 \\
CLJ0542.8-4100 &65.2 &1.57 &2.85 \\
A665 &20.0 &2.60 &0.676 \\
ZWCL1953 &32.6 &1.33 &0.0380 \\
MS0906.5+1110 &21.2 &0.0689 &0.0103 \\
A773 &47.5 &0.0561 &0.374 \\
A781 &25.4 &1.62 &5.61 \\
A963 &5.11 &0.367 &0.0163 \\
ZWCL3146 &4.46 &0.0888 &0.0116 \\
MS1054.5-0321 &151. &10.5 &3.48\\
CLJ1113.1-2615 &7.01 &2.51 &1.04 \\
V1121.0+2327 &49.4 &16.1 &0.907 \\
MS1137.5+6625 &6.73 &0.580 &0.00815 \\
A1413 &58.3 &0.119 &0.284 \\
V1221.4+4918 &50.8 &2.18 &1.76 \\
CLJ1226.9+3332 &1.90 &1.56 &0.0840 \\
A1758 &188. &1.12 &1.65 \\
RXJ1350.0+6007 &75.0 &20.9 &0.324 \\
MS1358.4+6245 &13.3 &0.286 &0.0335 \\
A1914 &15.9 &0.765 &0.00769 \\
A2034 &14.1 &0.495 &0.180 \\
RXJ1532.9+3021 &5.05 &0.00623 &0.0254 \\
A2111 &113. &0.685 &0.590 \\
A2218 &27.6 &0.281 &0.0556 \\
A2219 &142. &0.716 &1.03 \\
RXJ1716.9+6708 &42.3 &0.857 &0.870 \\
RXJ1720.1+2638 &5.49 &0.176 &0.0391 \\
A2261 &4.94 &0.308 &0.0966 \\
MS2053.7-0449 &15.8 &3.34 &0.525 \\
RXJ2129.6+0005 &18.0 &0.0227 &0.152 \\
MS2137.3-2353 &1.87 &0.0269 &0.0140 \\
A2390 &58.2 &0.344 &0.289 \\
CLJ2302.8+0844 &9.11 &4.65 &0.303 \\
\enddata
\end{deluxetable}

\begin{deluxetable}{l@{\hspace{5pt}}c@{\hspace{5pt}}c@{\hspace{5pt}}c}
\tabletypesize{\footnotesize}
\tablecaption{ Power Ratio Errors }
\tablewidth{0pt}
\tablecolumns{4}
\tablehead{
\colhead{ Cluster } & \colhead{$P_2/P_0$ ($\times10^{-7}$)} & 
\colhead{$P_3/P_0$ ($\times10^{-7}$)} & \colhead{$P_4/P_0$ ($\times10^{-7}$)}
}
\startdata

MS0015&37-55 [+0.71,-0.36]&0.098-1.2 [+0.0045,-0]&0.0013-0.15 [+6.5e-4,-3.8e-4] \\

CLJ0152&130-480 [+49,-21]&4.5-37 [+2.4,-1.0]&0.88-12 [+0.99,-0.47] \\

A267&43-79 [+1.8,-0.89]&0.020-1.0 [+0.0072,-0.0041]&0.055-0.94 [+0.0089,-0.0041] \\

MS0302&7.0-71 [+3.1,-1.4]&0.054-7.1 [+0.23,-0.11]&0.10-3.5 [+0.15,-0.065] \\

RXJ0439+07&27-60 [+1.3,-0.75]&0.22-3.7 [+0.025,-0.057]&0.011-0.70 [+0,-0.0028] \\

RXJ0439+05&0.21-5.5 [+0.68,-0.55]&0.0073-1.2 [+0.013,-0.0062]&0.0027-0.27 [+0.0037,-0.0019]\\

A520&22-49 [+0.74,-0.36]&1.3-5.2 [+0.040,-0.019]&0.51-2.8 [+0.031,-0.015] \\

MS0451&52-80 [+1.9,-0.95]&1.7-3.9 [+0.13,-0.060]&0.13-0.73 [+0.014,-0.0057]\\

CLJ0542&38-110 [+8.2,-3.7]&0.10-6.3 [+0.26,-0.12]&0.59-7.0 [+0.49,-0.23] \\

A665&13-30 [+1.7,-1.3]&1.2-5.0 [+0.33,-0.28]&0.34-1.5 [+0.014,-0.0072] \\

ZWCL1953&23-48 [+1.3,-0.64]&0.14-2.3 [+0.019,-0.0088]&0.0062-0.48 [+0.0055,-0.0022] \\

MS0906&16-26 [+0.78,-0.37]&0.0041-0.31 [+0.0032,-8.1e-4]&0.0018-0.11 [+3.3e-4,-3.7e-4] \\

A773&36-63 [+1.0,-0.49]&0.020-1.0 [+0.025,-0.027]&0.15-1.2 [+0.021,-0.022] \\

A781&13-60 [+1.3,-0.64]&0.24-7.3 [+0.090,-0.041]&1.7-7.8 [+0.14,-0.080] \\

A963&3.3-6.5 [+0.22,-0.10]&0.23-0.73 [+0.022,-0.010]&0.0015-0.069 [+0.0020,-9.0e-4]\\

ZWCL3146&3.5-5.7 [+0.072,-0.044]&0.035-0.19 [+0.0098,-0.0087]&0.0013-0.040 [+0.0017,-0.0015]\\

MS1054&120-180 [+9.3,-4.3]&6.0-15 [+0.85,-0.37]&2.1-5.5 [+0.30,-0.14] \\

CLJ1113&0.70-39 [+6.3,-2.1]&0.70-20 [+3.3,-1.1]&0.090-4.8 [+0.58,-0.21] \\

V1121&24-88 [+11,-4.5]&7.6-33 [+3.5,-1.5]&0.15-5.1 [+0.46,-0.15] \\

MS1137&2.1-19 [+2.4,-1.0]&0.035-2.5 [+0.12,-0.044]&0.020-0.79 [+0.058,-0.021] \\

A1413&50-67 [+0.84,-0.42]&0.027-0.54 [+0.0033,-0.0020]&0.15-0.59 [+0.0050,-0.0021] \\

V1221&24-79 [+5.9,-2.6]&0.80-8.0 [+0.35,-0.15]&0.73-5.7 [+0.38,-0.16] \\

CLJ1226&0.53-14 [+0.37,-0.16]&0.41-5.5 [+0.11,-0.058]&0.018-1.2 [+0.026,-0.0081] \\

A1758&170-200 [+6.5,-3.1]&0.71-2.1 [+0.042,-0.025]&1.2-2.2 [+0.072,-0.035] \\

RXJ1350&39-230 [+24,-9.5]&3.7-51 [+5.1,-1.8]&0.16-13 [+1.4,-0.50] \\

MS1358&9.7-18 [+0.51,-0.24]&0.10-0.60 [+0.021,-0.010]&0.0035-0.16 [+0.0058,-0.0027] \\

A1914&12-21 [+0.29,-0.14]&0.37-1.5 [+0.015,-0.0068]&0.0012-0.11 [+0.0011,-8.1e-4] \\

A2034&12-17 [+3.2,-2.6]&0.24-0.84 [+0.11,-0.080]&0.095-0.30 [+0.052,-0.044] \\

RXJ1532&3.1-8.4 [+0.066,-0.034]&0.0031-0.14 [+4.9e-4,-1.7e-4]&0.0044-0.13 [+0.0014,-8.0e-4]\\

A2111&78-150 [+5.3,-2.6]&0.081-2.9 [+0.077,-0.047]&0.036-1.5 [+0.028,-0.014] \\

A2218&22-33 [+3.1,-2.6]&0.12-0.91 [+0.078,-0.059]&0.011-0.21 [+0.015,-0.011] \\

A2219&130-150 [+2.4,-1.2]&0.44-1.0 [+0.0088,-0.0038]&0.87-1.3 [+0.018,-0.0088] \\

RXJ1716&17-82 [+7.7,-3.4]&0.15-6.6 [+0.30,-0.12]&0.058-3.6 [+0.14,-0.062] \\

RXJ1720&3.2-7.8 [+0.36,-0.21]&0.053-0.66 [+0.066,-0.057]&0.0038-0.17 [+0.0086,-0.0059] \\

A2261&1.7-6.5 [+0.075,-0.035]&0.013-0.80 [+0.010,-0.0051]&0.020-0.34 [+0.0018,-8.9e-4] \\

MS2053&2.7-42 [+5.2,-2.1]&0.53-14 [+1.1,-0.47]&0.051-4.0 [+0.31,-0.11] \\

RXJ2129&13-24 [+0.29,-0.14]&0.0085-0.63 [+0.0050,-0.0023]&0.023-0.36 [+0.0031,-0.0017] \\

MS2137&1.2-2.8 [+0.024,-0.0095]&0.0038-0.12 [+0.0014,-6.7e-4]&0.0035-0.066 [+2.1e-4,-8.7e-5]\\

A2390&52-67 [+0.074,-0.038]&0.17-1.0 [+0.0079,-0.044]&0.13-0.59 [+0.0036,-0.0021] \\

CLJ2302&0.74-42 [+6.5,-2.3]&0.81-19 [+3.7,-1.2]&0.067-3.7 [+0.58,-0.20] \\
\enddata
\small
\tablecomments{ 90\% confidence intervals for the power ratios based on Monte Carlo simulations.  The average systematic errors from the normalization of the background are listed in brackets. }
\end{deluxetable}

\begin{deluxetable}{lcccccc}
\tablecaption{ Statistical Significance of Results }
\tablewidth{0pt}
\tablecolumns{7}
\tablehead{
\colhead{} & \colhead{Average} & \colhead{Average} & \colhead{Rank-Sum} &
\colhead{KS Prob.} & \colhead{Average} & \colhead{Average}\\
\colhead{} & \colhead{Low-z} & \colhead{High-z} & \colhead{Prob.} & &
\colhead{RS Prob.} & \colhead{KS Prob.}
}
\startdata
$P_2/P_0$ &3.92e-6 &6.16e-6 &0.10 &0.34 &- &- \\
$P_3/P_0$ &6.93e-8 &5.95e-7 &4.6e-5 &0.00064 &0.00036 &0.0023 \\
$P_4/P_0$ &5.20e-8 &1.30e-7 &0.025 &0.041 &0.0082 &0.037 \\
\enddata
\tablecomments{ Columns 1 and 2 give the average power ratios for the high and low-redshift samples, respectively.  Column 3 lists the probability from a rank-sum test that the low and high-redshift clusters have the same average power ratios.  The probability from a KS-test that the two samples have the same distribution is given in column 4.  The last two columns list the average rank-sum and KS probabilities for 1000 runs where the power ratios were randomly selected from the Monte Carlo simulations. }
\end{deluxetable}

\begin{deluxetable}{lccc}
\tablecaption{ Noise Corrected Power Ratios }
\tablewidth{0pt}
\tablecolumns{4}
\tablehead{
\colhead{ Cluster } & \colhead{$P_2/P_0$ ($\times10^{-7}$)} & 
\colhead{$P_3/P_0$ ($\times10^{-7}$)} & \colhead{$P_4/P_0$ ($\times10^{-7}$)}
}
\startdata
MS0015.9+1609 &46.4 &0.316 &-0.0459 \\
CLJ0152.7-1357 &264. &12.4 &4.20 \\
A267 &61.4 &-0.306 &0.254 \\
MS0302.7+1658 &36.9 &0.658 &0.0479 \\
RXJ0439.0+0715 &46.2 &0.961 &0.0191 \\
RXJ0439.0+0520 &1.29 &-0.0455 &-0.0509 \\
A520 &40.7 &2.43 &0.870 \\
MS0451.6-0305 &66.1 &2.19 &0.316 \\
CLJ0542.8-4100 &61.3 &0.332 &2.31 \\
A665 &19.4 &2.43 &0.603 \\
ZWCL1953 &31.7 &1.05 &-0.0848 \\
MS0906.5+1110 &21.0 &0.00166 &-0.0181 \\
A773 &46.9 &-0.125 &0.297 \\
A781 &21.7 &0.512 &5.14 \\
A963 &5.03 &0.342 &0.00607 \\
ZWCL3146 &4.42 &0.0780 &0.00715 \\
MS1054.5-0321 &150. &10.3 &3.37 \\
CLJ1113.1-2615 &-2.16 &-0.482 &-0.302 \\
V1121.0+2327 &45.0 &14.7 &0.322 \\
MS1137.5+6625 &5.24 &0.115 &-0.195 \\
A1413 &58.0 &0.0525 &0.256 \\
V1221.4+4918 &47.9 &1.29 &1.37 \\
CLJ1226.9+3332 &-0.821 &0.770 &-0.243 \\
A1758 &188. &1.06 &1.62 \\
RXJ1350.0+6007 &58.8 &15.7 &-1.98 \\
MS1358.4+6245 &13.2 &0.228 &0.00851 \\
A1914 &15.7 &0.709 &-0.0157 \\
A2034 &14.1 &0.476 &0.172 \\
RXJ1532.9+3021 &4.89 &-0.0389 &0.00664 \\
A2111 &112. &0.210 &0.386 \\
A2218 &27.4 &0.233 &0.0353 \\
A2219 &142. &0.700 &1.02 \\
RXJ1716.9+6708 &37.8 &-0.541 &0.266 \\
RXJ1720.1+2638 &5.32 &0.124 &0.0175 \\
A2261 &4.57 &0.201 &0.0516 \\
MS2053.7-0449 &12.1 &2.16 &0.00641 \\
RXJ2129.6+0005 &17.6 &-0.0814 &0.108 \\
MS2137.3-2353 &1.81 &0.00772 &0.00588 \\
A2390 &58.0 &0.291 &0.267 \\
CLJ2302.8+0844 &1.41 &2.21 &-0.767 \\
\enddata
\end{deluxetable}

\begin{deluxetable}{lccccc}
\tablecaption{ Results After Noise Correction}
\tablewidth{0pt}
\tablecolumns{5}
\tablehead{
\colhead{} & \colhead{Average} & \colhead{Average} & \colhead{Rank-Sum} &
\colhead{KS Prob.} & \colhead{Slope}\\
\colhead{} & \colhead{Low-z} & \colhead{High-z} & \colhead{Prob.} & & \colhead{vs. z}
}
\startdata
$P_2/P_0$ &3.84e-6 &5.67e-6 &0.25 &0.34 &- \\
$P_3/P_0$ &4.67e-8 &4.39e-7 &0.013 &0.015 &4.09e-7 (8.19e-8, 8.03e-7) \\
$P_4/P_0$ &4.25e-8 &6.17e-8 &0.31 &0.15 &7.47e-8 (-2.57e-8, 2.22e-7) \\
\enddata
\tablecomments{ Columns 1 and 2 give the average power ratios for the high and low-redshift samples after the noise correction.  Column 3 lists the probability from a rank-sum test that the low and high-redshift clusters have the same average power ratios.  The probability from a KS-test that the two samples have the same distribution is given in column 4.  The last column lists the average slope of the power ratios vs. z from 1000 fits using a least absolute deviation method and where the power ratios were randomly selected from the Monte Carlo simulations. The 90\% confidence limits on the slope are given in parentheses. }
\end{deluxetable}

\end{document}